\documentclass[12pt,preprint]{aastex}
\usepackage{epsfig}
\usepackage{natbib}
\usepackage{graphicx}
\usepackage{slashbox}
\usepackage{multirow}
\usepackage{lscape}
\usepackage{mathrsfs,amssymb}
\usepackage{amsmath}
\usepackage{subfigure}
\usepackage{amssymb}
\usepackage{multirow}
\usepackage{tabularx}
\usepackage{rotating}
\usepackage{amsmath}
\usepackage{ulem}
\newcommand       \cm           {\,{\rm cm}}

\newcommand       \K            {\,{\rm K}}

\newcommand       \NH           {N_{\rm H}}
\newcommand       \simlt        {\lesssim}
\newcommand       \simgt        {\gtrsim}

\newcommand       \mum          {\,{\rm \mu m}}
\newcommand       \ppm          {\,{\rm ppm}}

\newcommand       \simali       {\sim\,}

\newcommand \dprim {\left[\rm D/H\right]_{\rm prim}}

\newcommand \dism {\left[\rm D/H\right]_{\rm ISM}}

\newcommand \dgas {\left[\rm D/H\right]_{\rm gas}}

\newcommand \dpah {\left[\rm D/H\right]_{\scriptscriptstyle\rm PAH}}
\newcommand \cpah {\left[\rm C/H\right]_{\scriptscriptstyle\rm PAH}}

\newcommand       \Acd           {A_{4.65}}
\newcommand       \Aaro          {A_{3.3}}
\newcommand       \Aali          {A_{3.4}}

\newcommand       \Aratio        {A_{4.65}/A_{3.3}}
\newcommand       \ACD           {A_{4.4}}
\newcommand       \ACH           {A_{3.3}}
\newcommand \faliCH {f_{\rm aliCH}}
\newcommand \faliCD {f_{\rm aliCD}}
\newcommand \faroCD {f_{\rm aroCD}}
\newcommand       \NC         {N_{\rm C}}
\newcommand       \ND         {{N_{\rm D}}}
\newcommand       \Iratioobs     {\left(I_{4.65}/I_{3.3}\right)_{\rm obs}}

\newcommand       \NCH         {N_{\rm C-H}}
\newcommand       \NCD         {{N_{\rm C-D}}}
\newcommand       \km        {\,{\rm km}}

\newcommand       \mol       {\,{\rm mol}}

\newcommand       \tauch       {\Delta\tau_{3.4\mum}}
\newcommand       \taucd       {\Delta\tau_{4.65\mum}}
\countdef\decade=200
\advance\decade by \year
\countdef\hours=201
\advance\hours by \time
\divide\hours by 60
\countdef\mins=202
\advance\mins by \hours
\multiply\mins by 60
\multiply\hours by 100
\countdef\miltime=203
\advance\miltime by \hours
\advance\miltime by \time
\advance\miltime by -\mins
\def\today{\number\decade.\number\month.\number\day.\number\miltime}
\shorttitle{IR spectra of Deuterated PAHs}
\title{
Deuterated Polycyclic Aromatic Hydrocarbons
in the Interstellar Medium:
The Aliphatic C--D Band Strengths
\\{\small DRAFT: \today ~~}
}
\author{X.J.~Yang\altaffilmark{1,2}
            and Aigen Li\altaffilmark{2}
            }
\altaffiltext{1}{Department of Physics,
                      Xiangtan University,
                      411105 Xiangtan, Hunan Province, China;
                      {\sf xjyang@xtu.edu.cn}}
\altaffiltext{2} {Department of Physics and Astronomy,
                  University of Missouri,
                  Columbia, MO 65211, USA;
                  {\sf lia@missouri.edu}}

\begin{document}

\begin{abstract}
Deuterium (D) was exclusively generated in the Big Bang
and the standard Big Bang Nucleosynthesis (BBN) model
predicts a primordial abundance of D/H\,$\approx$\,26
parts per million (ppm). As the Galaxy evolves,
D/H gradually decreases because of astration.
The Galactic chemical evolution (GCE) model predicts
a present-day abundance of D/H\,$\simgt$\,20$\ppm$.
However, observations of the local interstellar medium
(ISM) have revealed that the gas-phase interstellar D/H
varies considerably from one region to another
and has a median abundance of D/H\,$\approx$\,13$\ppm$,
substantially lower than predicted from
the BBN and GCE models.
It has been suggested that the missing D atoms
of D/H\,$\approx$\,7$\ppm$ could have been
locked up in deuterated polycyclic aromatic
hydrocarbon (PAH) molecules.
%However, by analyzing the aromatic C--D
%stretching emission from deuterated PAHs
However, we have previously demonstrated
that PAHs with aromatic C--D units
are insufficient to account for the missing D.
Here we explore if PAHs with aliphatic C--D
units could be a reservoir of D.
We perform quantum chemical computations
of the vibrational spectra of ``superdeuterated''
PAHs (in which one D and one H share an C atom)
and PAHs attached with D-substituted methyl group,
and derive the band strengths of
the aliphatic C--D stretch ($\Acd$).
By applying the computationally derived $\Acd$
to the observed aliphatic C--D emission
at $\simali$4.6--4.8$\mum$,
we find that PAHs with aliphatic C--D units
could have tied up a substantial amount of D/H
and marginally account for the missing D.
The possible routes to generate PAHs with
aliphatic C--D units are also discussed.
\end{abstract}

\keywords {dust, extinction --- ISM: lines and bands
           --- ISM: molecules}

%Polycyclic aromatic hydrocarbons (1280); Interstellar line emission
%(844); Line intensities (2084); Astrochemistry (75); Interstellar
%molecules (849)

\section{Introduction\label{sec:intro}}
A set of infrared (IR) emission features
at 3.3, 6.2, 7.7, 8.6, and 11.3$\mum$,
commonly attributed to polycyclic aromatic
hydrocarbon (PAH) molecules,
are ubiquitously seen in a wide variety
of astrophysical environments.
We should emphasize that astronomical
PAHs are not necessarily
{\it pure} aromatic compounds
as strictly defined by chemists.
Instead, PAH molecules in space
may include ring defects, substituents,
partial dehydrogenation, or superhydrogenation
(see Yang et al.\ 2017 and references therein).
Astronomical PAHs could also be deuterated
(e.g., see Allamandola et al.\ 1989;
Hudgins et al.\ 2004; Peeters et al.\ 2004;
Draine 2006; Onaka et al.\ 2014, 2022;
Mori et al.\ 2022; Yang et al.\ 2020, 2021).
Indeed, it has been suggested that PAHs
could be a reservoir of deuterium (D)
in the interstellar medium (ISM; Draine 2006).
% and account for the D atoms missing from
%the gas phase (Draine 2006; see below).

According to the standard Big Bang
Nucleosynthesis (BBN) model, D was
exclusively created in the Big Bang
with a primordial D/H abundance of
$\dprim$\,$\approx$\,26$\ppm$
(see Yang et al.\ 2020 and references therein).
Due to astration (i.e., D is converted to
$^3$He, $^4$He, and heavier elements
in stellar interiors),
the present-day Galactic interstellar D/H
abundance should be reduced
from the primordial abundance
to $\dism$\,$\simgt$\,$20\ppm$.
However, the gas-phase $\dgas$ abundance
of the local ISM derived from high resolution
ultraviolet (UV) spectroscopic observations
varies considerably from one region to another,
and ranges from $\simali$5$\ppm$
to $\simali$22$\ppm$, with a median abundance
of $\dgas\approx13\ppm$ (see Draine 2006).
This implies that there must exist a sink
to accommodate the missing D/H,
typically of $\simali$7$\ppm$.
%
%
%...and in some regions it is substantially lower
%than the expected interstellar abundance of
%$\dism$\,$\simgt$\,$20\ppm$
%(e.g., $\dgas$\,=\,$5.0\pm1.6\ppm$
%toward $\theta$~Car;
%see Draine 2006 and referenes therein).
%
%
The exact sink for the missing D atoms
remains unknown. If PAHs are responsible
for the missing D atoms, they should
typically have tied up an amount of
$\dpah\approx7\ppm$.
Here $\dpah$ refers to the amount
of D locked up in PAHs
(relative to the total H in the ISM).

%Deuteration could occur
%during photodissociation events.
%Upon absorption of an energetic photon,
%one or more periphal hydrogen (H) atoms
%could be stripped off and replaced by D atoms.
%Due to the zero-point energy difference,
%C--D bonds are more stable than C--H bonds
%and therefore PAHs of intermediate size
%are expected to become deuterium enriched
%in the ISM.

If interstellar PAHs are deuterated through replacing
one or more of their periphal hydrogen (H) atoms
by D atoms, one expects to see an emission band
at $\simali$4.4$\mum$, arising from the aromatic
C--D stretch, in analogy to the 3.3$\mum$ C--H stretch.
Indeed, the 4.4$\mum$ C--D stretching emission band
has been detected in photodissociated regions (PDRs),
reflection nebulae, and H\,{\sc ii} regions
both in the Milky Way and in the Large and Small
Magellanic Clouds (Peeters et al.\ 2004;
Onaka et al.\ 2014, 2022; Doney et al.\ 2016).
The intrinsic band strengths of the 4.4$\mum$
C--D stretch have been computed
by Yang et al.\ (2020, 2021)
and Buragohain et al.\ (2015, 2016, 2020)
and have been applied to astronomical observations
to estimate the degree of D enrichment in PAHs.
Yang et al.\ (2020, 2021) have found that
small interstellar PAHs could indeed be D-enriched,
exceeding the interstellar D/H abundance
by a factor of $\simgt$1000.
Nevertheless, this is still not sufficient
to account for the missing D atoms.
%(see \S\ref{sec:astro}).
%

As will be discussed in \S\ref{sec:routes},
PAHs could also be deuterated by attaching
D-containing aliphatic sidegroups
(e.g., --CH$_2$D)
to their edge carbon (C) atoms,
with the asymmetric and symmetric
aliphatic C--D stretching modes
occuring at $\simali$4.6--4.8$\mum$
(see Tielens 1997).
Indeed, these C--D emission bands
have also been seen in interstellar sources,
together with the 4.4$\mum$ aromatic
C--D stretch (Peeters et al.\ 2004;
Onaka et al.\ 2014, 2022; Doney et al.\ 2016).
In regions rich in H and D atoms, PAHs could
be superhydrogenated by inserting an extra
H atoms to an edge C atom so that two H atoms
share one C atom. This converts aromatic C--H
stretchs to aliphatic. In analogy, an D atom
could also be squeezed in the periphal, originally
aromatic C--H  unit to create an aliphatic C--H unit
and an aliphatic C--D unit.
%In dense regions where dust grains are coated by
%an ice mantle, upon condensation onto
%the ice mantle and exposed to UV irradiation,
%PAHs could be deuterated if interstellar ice
%is appreciably deuterated.

The goal of this work is to evaluate the role of PAHs
containing aliphatic C--D units as a potential reservoir of
D atoms. To this end, we first perform quantum chemical
computations of the vibrational spectra of PAHs with
deuterated aliphatic sidegroups and ``superdeuterated''
PAHs in which one H atom and one D atom share one C atom.
The computational methods and target PAH molecules
are summarized in \S\ref{sec:Method}.
The computed spectra as well as the aliphatic C--D band
strengths derived from the computed spectra are presented
in \S\ref{sec:results}. In \S\ref{sec:Obs} we compile all
the astronomical data for the aliphatic C--D stretch.
In \S\ref{sec:dpah} we apply the aliphatic C--D band strengths
derived here to astronomical observations.
We quantitatively determine the amount of D atoms
possibly locked up in PAHs with aliphatic C--D units.
The possible routes for aliphatically deuterating
PAHs in the ISM are discussed in \S\ref{sec:routes}.
Finally, we summarize our major results
in \S\ref{sec:summary}.
This paper is largely concerned with
the quantum chemical computations
of PAHs containing aliphatic C--D units.
Readers who are interested only in
the aliphatic C--D band intensities
and the astrophysical applications
may wish to proceed directly to
\S\ref{sec:BandStrengths}.

%This paper is organized as follows.
%In \S\ref{sec:Method} we briefly describe
%the computational methods and
%the selected target molecules.
%The computed IR vibrational spectra and
%the derived intrinsic C--D band strengths
%are reported in \S\ref{sec:Results}.
%In \S\ref{sec:astro} we apply
%the derived band strengths to
%to determine the deuteration degrees
%of interstellar PAHs, and discussion in
%\S\ref{sec:discussion}.
%Finally, we summarize our major results
%in \S\ref{sec:summary}.
%

\section{Computational Methods
         and Target Molecules
         \label{sec:Method}
         }
We use the Gaussian16 software (Frisch et al.\ 2016)
to calculate the IR vibrational spectra for different
parent PAH molecules containing aliphatic C--D units.
We consider seven parent aromatic molecules,
i.e., benzene (C$_6$H$_6$),
naphthalene (C$_{10}$H$_8$),
anthracene (C$_{14}$H$_{10}$),
phenanthrene (C$_{14}$H$_{10}$),
pyrene (C$_{16}$H$_{10}$),
perylene (C$_{20}$H$_{12}$), and
coronene (C$_{24}$H$_{12}$).
Here we focus on small PAHs with fewer than
24 C atoms (i.e., $\NC\simlt24$)
since deuterium enrichment in PAHs
of $\NC\simgt40$ is not expected.
Large PAHs have a sufficiently large
number of internal degrees of freedom
to accommodate the maximum energy
of typical UV photons without subsequent
photolytic bond cleavage occurring
(Hudgins et al.\ 2004).
Nevertheless, if the deuteration of
PAHs occurs in D-enriched ice mantles
through D-atom addition (see below),
large PAHs could also be deuterated.

\begin{figure*}
\centering{
\includegraphics[scale=0.5,clip]{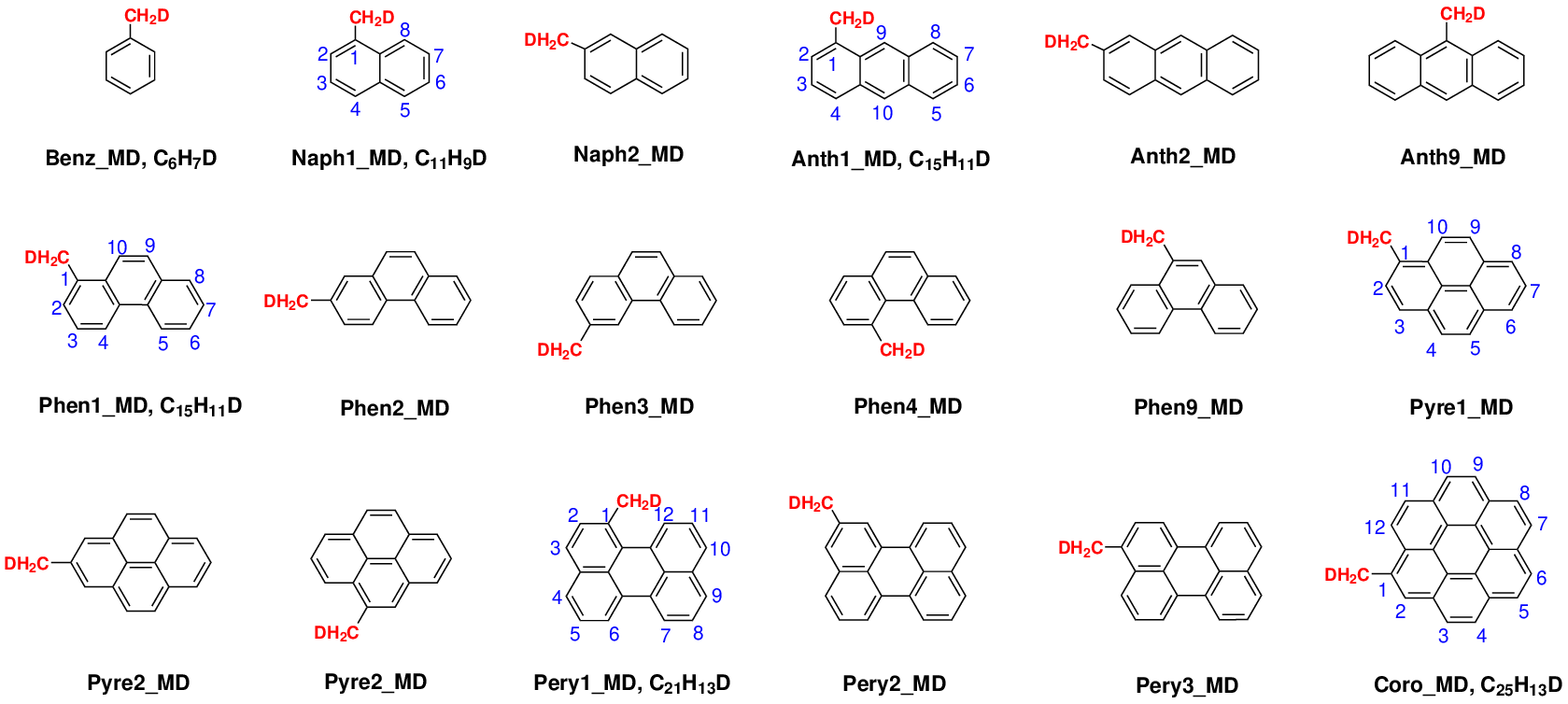}
}
\caption{\footnotesize
         \label{fig:PAH_MethylD_Scheme}
         Structures of methyl-deuterated PAHs.
         We refer to a methyl-deuterated species
         by the abbreviation of the first four letters of
         the name of its parent molecule
         followed by the position where the methyl group
         is attached (e.g., Anth2 refers to anthracene
         with the deuterated methyl attached at position 2).
         Since there are three H atoms in methyl and any
         of them could be deuterated,
         therefore all three possibilities
         are taken into account.
         }
\end{figure*}

\begin{figure*}
\centering{
\includegraphics[scale=0.45,clip]{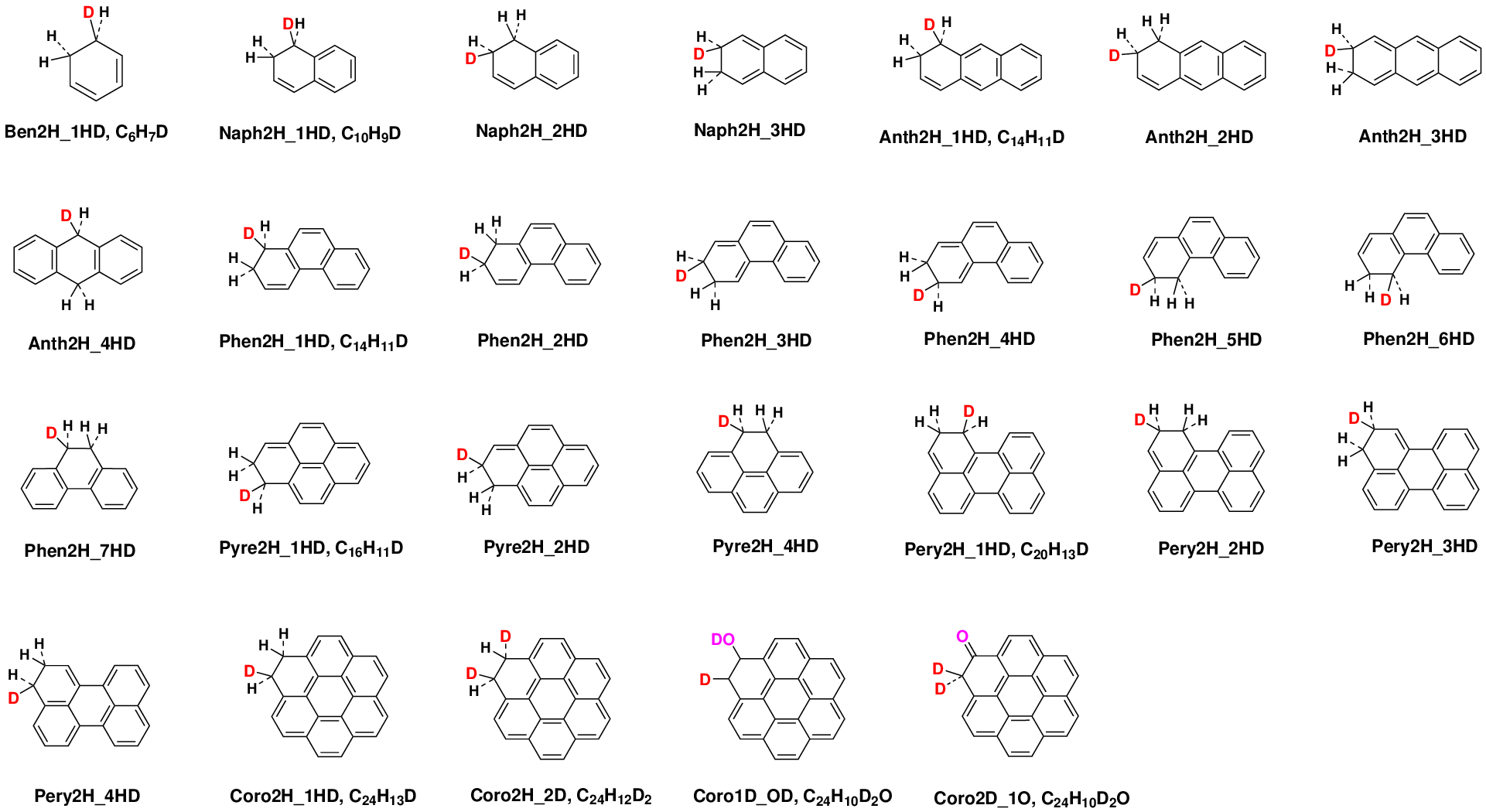}
}
\caption{\footnotesize
         \label{fig:PAH_HD_Scheme}
         Structures of ``superdeuterated'' PAHs
         in which one D and one H atom share an C atom.
         Since H is much more abundant than D in space,
         if a PAH molecule is ``superdeuterated'',
         it is very likely that it will also be superhydrogenated.
         Also shown are doubly ``superdeuterated'' coronene
         (with two pairs of H and D atoms),
         and deuterated, oxidized coronenes
         (see \S\ref{sec:routes}).
         }
\end{figure*}

We consider PAHs with two categories of
aliphatic C--D units,
i.e., PAHs attached with one deuterated
methyl group (hereafter PAH$\_$CH$_{2}$D;
see Figure~\ref{fig:PAH_MethylD_Scheme})
and ``superdeuterated'' PAHs
(hereafter PAH$\_$HD;
see Figure~\ref{fig:PAH_HD_Scheme})
%in which one C atom is attached with two H atoms
%and a neighbouring C atom is attached
%with both H and D atoms.
in which one edge C atom is shared by
one H atom and one D atom.
For both categories, various isomers are considered.
As H is much more abundant than D,
if a PAH molecule is ``superdeuterated'',
it is very likely that it will also be superhydrogenated
(i.e., two H atoms share one C atom).
Therefore, we consider ``superdeuterated'' PAHs
to be superhydrogenated as well.

In dense regions where dust grains are coated by
an ice mantle, upon condensation onto
the ice mantle and exposed to UV irradiation,
PAHs could be deuterated if interstellar ice
is appreciably deuterated.
Laboratory studies have shown that
UV photolysis of PAHs in D-enriched ices
results in rapid D enrichment of PAHs
through aromatic D$\rightarrow$H exchange,
D-atom addition (``superdeuteration''),
and exchange through keto-enol tautomerism
(see Figure~4 of Sandford et al.\ 2000).
The latter two reactions could also occur
in the absence of UV irradiation.
For illustrative purpose, we therefore also
consider a deuterated molecule with
a ketone (C=O) bond and a deuterated
molecule with a deuteroxyl (OD) sidegroup,
using coronene as an example
(see Figure~\ref{fig:PAH_HD_Scheme}).

\begin{figure*}
\centering{
\includegraphics[scale=0.5,clip]{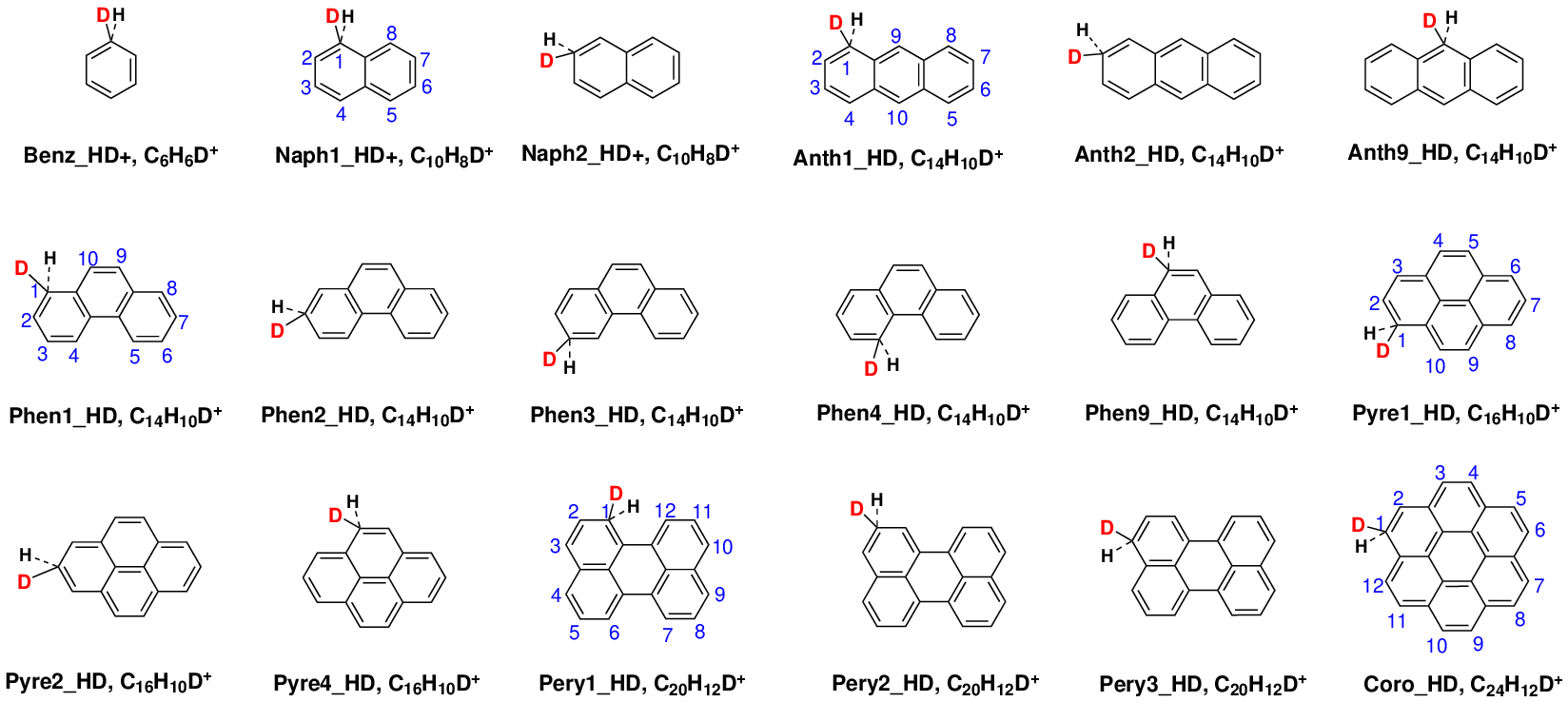}
}
\caption{\footnotesize
         \label{fig:PAH_HDPlus_Scheme}
         Structures of ``superdeuterated'' PAH cations.
         }
\end{figure*}

To examine the effects of ionization on
the vibrational spectra and aliphatic C--D
band strengths of deuterated PAHs
containing aliphatic C--D units,
we also consider the cationic counterparts
of ``superdeuterated'' PAHs (i.e., PAH$\_$HD).
For simplicity, we only consider PAHs
with a single H and D pair
(see Figure~\ref{fig:PAH_HDPlus_Scheme}).
They are essentially the radical counterparts
of those molecules shown in
Figure~\ref{fig:PAH_HD_Scheme},
with one H atom stripped off
from their H and H pair.

\begin{figure*}
\centering{
\includegraphics[scale=0.45,clip]{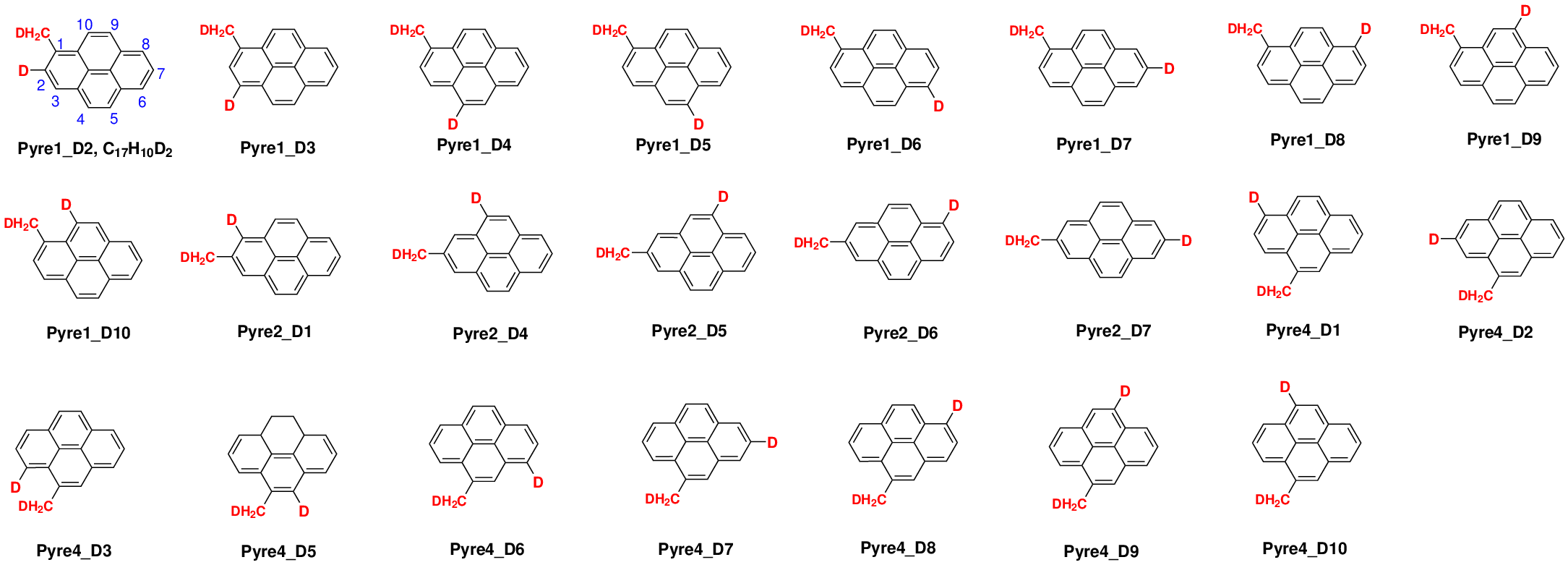}
}
\caption{\footnotesize
         \label{fig:PAH_Methyl2D_Scheme}
         Structures of 23 isomers of deuterated pyrenes
         containing a methyl-deuterated sidegroup
         as well as a periphal D atom.
         These molecules have both aliphatic
         and aromatic C--D bonds.
         }
\end{figure*}

\begin{figure*}
\centering{
\includegraphics[scale=0.45,clip]{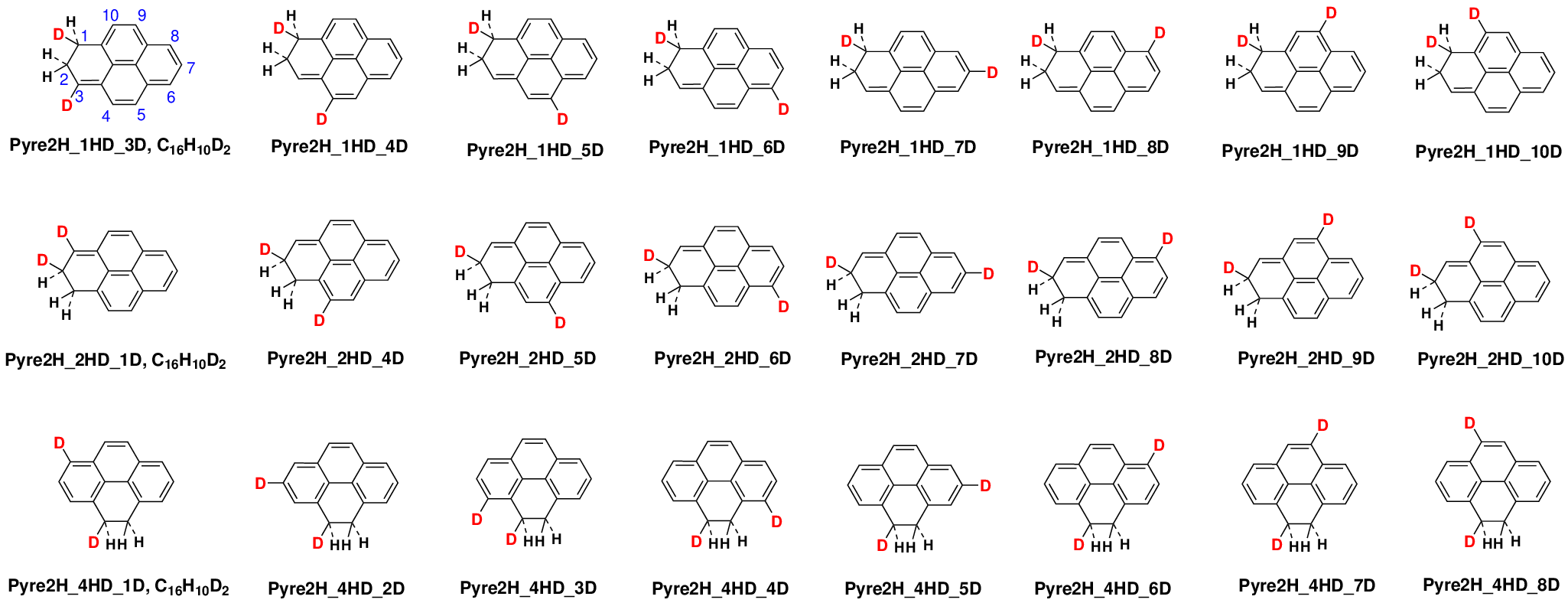}
}
\caption{\footnotesize
         \label{fig:PAH_H2D_Scheme}
         Structures of 24 isomers of ``superdeuterated'' pyrenes
         containing an H+H pair, an H+D pair,
         as well as a periphal D atom.
         Similar to that in Figure~\ref{fig:PAH_Methyl2D_Scheme},
         these molecules also have both aliphatic
         and aromatic C--D bonds.
          }
\end{figure*}

In view of the fact that, observationally, the aliphatic
C--D emission at $\simali$4.6--4.8$\mum$ is
always seen together with the aromatic C--D emission
at $\simali$4.4$\mum$ (Peeters et al.\ 2004;
Onaka et al.\ 2014, 2022; Doney et al.\ 2016),
we also consider PAHs with both aromatic
and aliphatic C--D bonds,
taking pyrene as an example.
We consider pyrenes in which
an D atom substitutes an H atom
{\it and} a deuterated methyl group
(--CH$_2$D) is also attached
(hereafter PyreD$\_$CH$_{2}$D;
see Figure~\ref{fig:PAH_Methyl2D_Scheme}).
We also consider mono-deuterated pyrenes
which are ``superdeuterated'' as well
(hereafter PyreD$\_$HD;
see Figure~\ref{fig:PAH_H2D_Scheme}).
Again, as D is much less abundant than H,
``superdeuterated'' pyrenes are very likely
also superhydrogenated.
Different isomers are considered,
with D atoms and functional groups
attached at different positions.

We will refer PAHs with deuterated methyl sidegroups
(see Figure~\ref{fig:PAH_MethylD_Scheme})
by the abbreviation of the first four letters of
the names of their parental PAH molecules,
followed by the position where the methyl group
is attached and then ``MD''
(e.g., Naph1$\_$MD refers to naphthalene
with a deuterated-methyl group attached at position 1).
For ``superdeuterated'' PAHs
with an H and D pair as well as an H and H pair
(see Figure~\ref{fig:PAH_HD_Scheme}),
we use the abbreviation as following:
the first four letters of the names of
their parental molecules followed
by ``2H'' and then the order of the isomer
followed by ``HD'' (e.g., Naph2H$\_$2HD
refers to the second isomer of
superhydrogenated, superdeuterated naphthalene).
For ``superdeuterated'' PAH cations
(see Figure~\ref{fig:PAH_HDPlus_Scheme}),
we will also refer them by the abbreviation
of the first four letters of the names of their
parental molecules,
followed by the position where the D atom
is inserted and then ``HD+''
(e.g., Naph1$\_$HD+ refers to
naphthalene cation with the H and D pair
attached at position 1).
%
% For coronene derivatives,
% we start with ``Coro'' followed by
%the number of  D atoms and then ...
%
For deuterated pyrenes attached with CH$_{2}$D
(see Figure~\ref{fig:PAH_Methyl2D_Scheme}),
we use ``Pyre'' followed by the position
where the methyl group
is attached and then a letter of ``D'' followed by
the position of the aromatic C--D bonds
(e.g., Pyre1$\_$D3 refers to pyrene with
the deuterated-methyl group attached at position 1
and another D atom attached at position 3).
For deuterated pyrenes which are also ``superdeuterated''
(see Figure~\ref{fig:PAH_H2D_Scheme}),
we use ``Pyre2H'' followed by the position
where the H and D pair is attached,
followed by ``HD'' and then ``D''
which is further followed by
the position of the aromatic C--D bonds
(e.g., Pyre2H$\_$1HD$\_$D3 refers to
superdeuterated pyrene with the H and D pair attached
at position 1 and another D atom attached at position 3).

We employ the hybrid density functional theoretical
method (B3LYP) at the {\rm 6-311+G$^{\ast\ast}$} level,
which gives sufficient calculational accuracies
with operable computer time
(see Yang et al.\ 2017 and reference therein).
The standard scaling is applied to
the frequencies by employing
a scale factor of $\simali$0.9688
(Borowski 2012).
We should note that the aromatic C--D band
strengths of deuterated PAHs
derived by Yang et al.\ (2020, 2021)
from B3LYP/{\rm 6-311+G$^{\ast\ast}$} computations
are consistent with recent laboratory measurements
of deuterated quenched carbonaceous composites
(Mori et al.\ 2022).\footnote{%
  The DFT-computed intensity ratio
    of the 4.4$\mum$ aromatic C--D stretch
    to the 3.3$\mum$ aromatic C--H stretch
    is $\approx0.56\pm0.19$
    for mono-deuterated PAHs (Yang et al.\ 2020)
    and $\approx0.56\pm0.03$
    for multi-deuterated PAHs (Yang et al.\ 2021),
    while the experimentally measured ratio
    is $\approx0.56\pm0.04$ (Mori et al.\ 2022).
  }

\section{Vibrational Spectra and Aliphatic C--D
            Band Strengths of Deuterated PAHs
            \label{sec:results}}
\subsection{PAHs Attached with a D-substituted
                    Methyl Group
                    \label{subsec:MethylPAH_Results}}
For PAHs attached with a D-substituted methyl group
(i.e., PAH$\_$CH$_{2}$D), we calculate the vibrational
spectra of 18 isomers of seven parent PAH molecules
(see Figure~\ref{fig:PAH_MethylD_Scheme}),
including all the possible positions
where the methyl group is attached.
Since each of the three H atoms
in the methyl group (--CH$_3$)
could be substituted with an D atom,
therefore in total we consider 54 isomers.

\begin{figure*}
\centering{
\includegraphics[scale=0.4,clip]{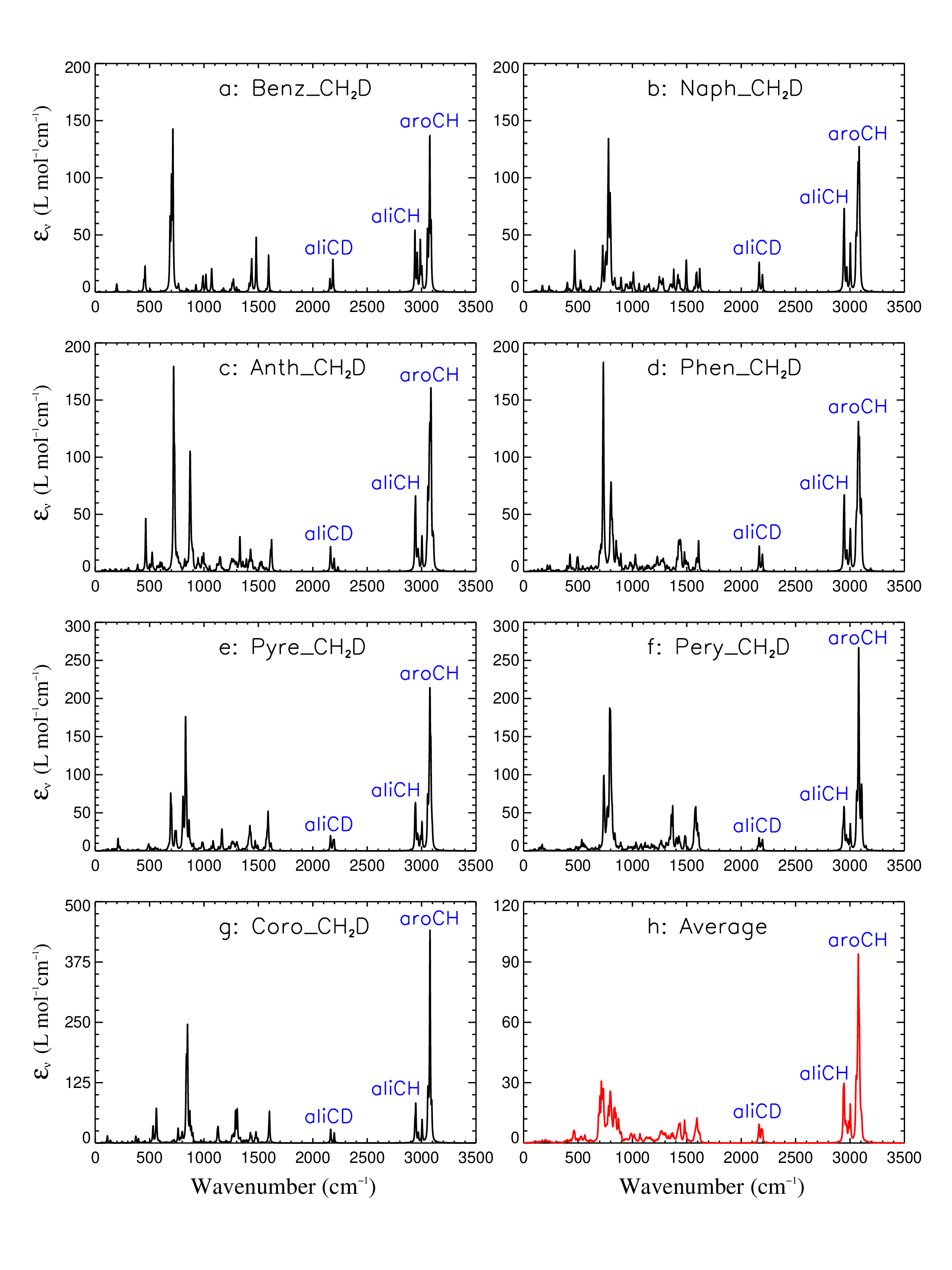}
}
\caption{\footnotesize
         \label{fig:PAH_Methyl1D_Spec}
         DFT-computed vibrational spectra
         of methyl-deuterated PAHs
         (see Figure~\ref{fig:PAH_MethylD_Scheme}),
         obtained by averaging over
         all the isomers of each parent molecule.
         Also shown is the overall mean spectrum
         obtained by averaging over the mean spectra
         of all seven parent molecules
         (normalized by $N_{\rm C}$).
         The aliphatic C--D, aliphatic C--H,
         and aromatic C--H stretches
         are labelled respectively by
         ``aliCD'', ``aliCH'' and ``aroCH''.
         }
\end{figure*}

For each parent molecule, we obtain
the mean spectrum averaged over
all the isomers and show in
Figures~\ref{fig:PAH_Methyl1D_Spec}a--g.
For each vibrational transition,
we assign a width of 4$\cm^{-1}$.
%and the vibrational intensities
%are given on a per-bond-basis in this paper,
%unless otherwise stated.
A feature at $\simali$2170$\cm^{-1}$
(i.e., $\simali$4.61$\mum$), originating
from aliphatic C--D stretches,
is clearly seen in all these mean spectra.
Also prominent are aliphatic and aromatic
C--H stretches at $\simali$2980$\cm^{-1}$
(i.e., $\simali$3.36$\mum$)
and 3060~$\cm^{-1}$
(i.e., $\simali$3.30$\mum$), respectively.
A close inspection of the aliphatic C--D stretches
shown in Figures~\ref{fig:PAH_Methyl1D_Spec}a--g
reveals two or three peaks, attributed to different
isomers. Nevertheless, these peaks are very close
to each other, indicating that the centroids of
the C--D stretches do not vary much among
different isomers.
As a matter of fact, the other vibrational modes
are not significantly blended either,
indicating that the transition frequencies for
different isomers and different molecules are
very similar, i.e, neither the position of the methyl
group attached nor the position of D atoms
in the methyl group affects the centroids of
the vibrational features much.

Figure~\ref{fig:PAH_Methyl1D_Spec}h
shows the overall mean spectrum
obtained by averaging over the mean spectra
of all seven parent molecules
(normalized by $N_{\rm C}$).
Again, the aliphatic C--D and C--H stretches
are clearly seen, while the aromatic C--H stretch
is the most prominent.
The aliphatic C--D stretch also narrowly peaks,
implying that the centroids of the C--D stretches
do not vary much among different molecules.
As the C--D in-plane and out-of-plane bending modes
are often well mixed with the C--H out-of-plane bending
modes and the C--C--C skeletal vibrational modes
and therefore difficult to identify (Yang et al.\ 2020, 2021),
we shall focus on the C--D and C--H stretching modes.

\begin{figure*}
\centering{
\includegraphics[scale=0.3,clip]{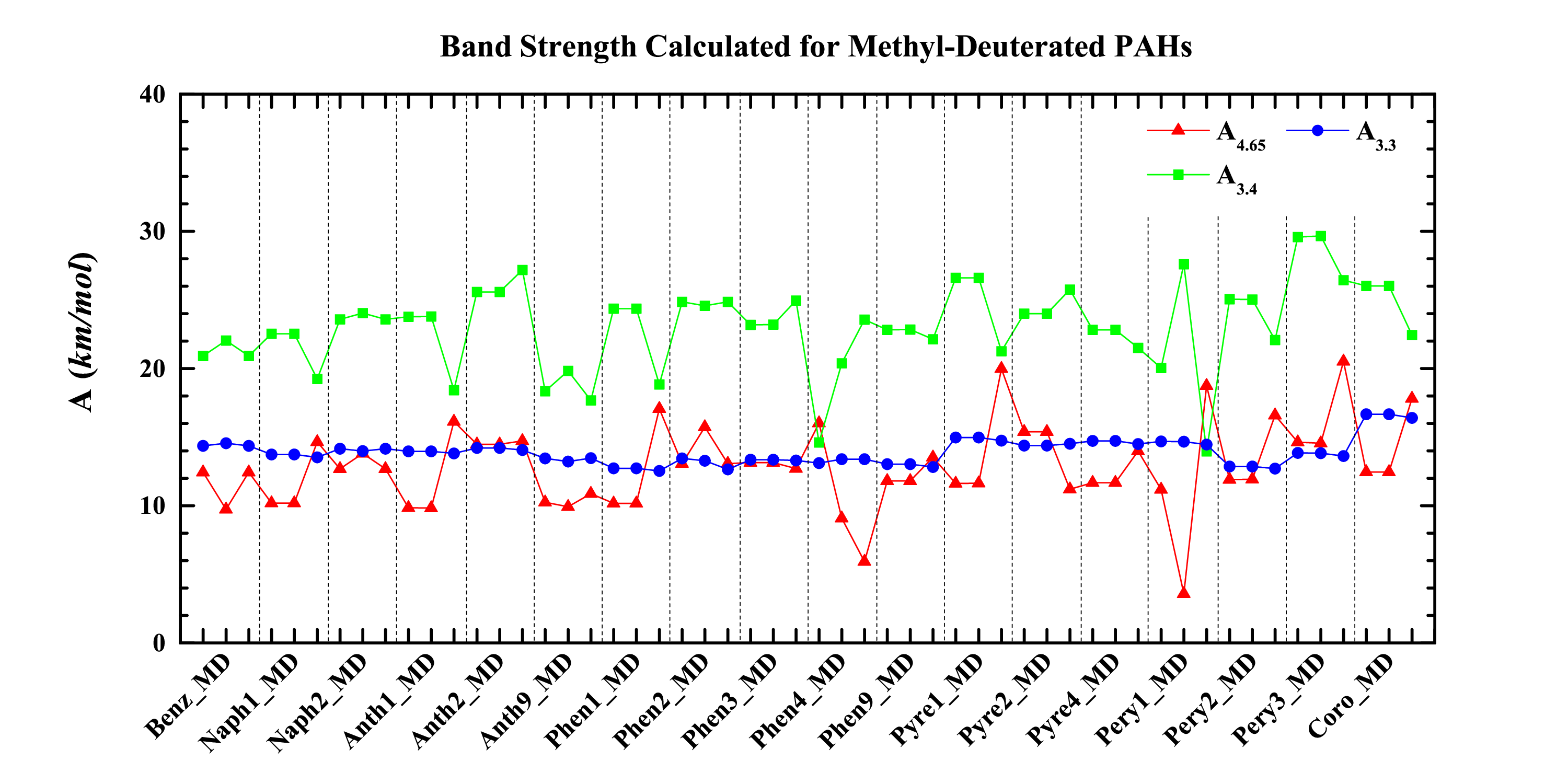}
}
\caption{\footnotesize
               \label{fig:Methyl1D_Intensity}
               Band strengths of the 4.65$\mum$
               aliphatic C--D stretches ($\Acd$; red),
               the 3.4$\mum$ aliphatic C--H stretches
               ($\Aali$; green), and the 3.3$\mum$
               aromatic C--H stretches
               ($\Aaro$; blue) computed
               at level {\rm B3LYP/6-311+G$^{\ast\ast}$}
               for methyl-deuterated PAHs.
               }
\end{figure*}

We tabulate the peak wavelengths and intensities
of the C--H and C--D stretch features in
Table~\ref{tab:Freq_Int_PAH_MethylD}.
The aliphatic C--D stretches generally occur at
$\simali$4.60$\mum$ and the deviation is negligible.
In Figure~\ref{fig:Methyl1D_Intensity} we show
the intensities of the aliphatic C--D stretches ($\Acd$),
the aliphatic C--H stretches ($\Aali$), and
the aromatic C--H stretches ($\Aaro$)
for all 54 isomers calculated here.
The intensities of these stretches,
especially the aromatic C--H stretches,
show small scatters among different isomers.
For aliphatic C--D stretches, the scatters
mostly arise from the different positions
of D atoms in the methyl group
attached to the parent molecule.
%For example, in Pery$\_$MM1,
%$\Acd$ varies in the range of 3--18$\km\mol^{-1}$.
As D atoms substitute H atoms in the methyl group
in a random manner, we should average the intensities
over that for the three positions of the D atoms
in the methyl group.
This should also be the case for the methyl group
attached at different positions of a parent molecule.
Nevertheless, the scatters in the average intensities
among the different isomers
of the same parent molecule are small
(see Table~\ref{tab:Freq_Int_PAH_MethylD}).
The overall mean band intensities,
on a per C--D or C--H bond basis, are
$\langle \Acd\rangle \approx 12.9\pm0.89\km\mol^{-1}$,
$\langle \Aali\rangle \approx 23.1\pm1.3\km\mol^{-1}$, and
$\langle \Aaro\rangle \approx 14.3\pm1.1\km\mol^{-1}$,
respectively for the aliphatic C--D stretch,
the aliphatic C--H stretch,
and the aromatic C--H stretch.

\subsection{``Superdeuterated'' PAHs
                    \label{subsec:PAH_HD_Results}}
The superhydrogenation of PAHs often starts from
the broken of a C=C bond, followed by the addition
of a pair of two H atoms in the neighbouring C atoms.
By ``superdeuterated'' PAHs (PAH$\_$HD),
we refer to superhydrogenated PAHs
but with one of the H atoms
in the H and H pair
substituted by an D atom,
not two D atoms share one C atom.
As mentioned earlier, H is much more
abundant than D in space,
if a PAH molecule is ``superdeuterated'',
it is very likely that it will also be superhydrogenated.
Therefore, for ``superdeuterated'' PAHs
we consider them to have a pair of two H atoms
sharing an C atom, and meanwhile a pair of
H and D atoms sharing another C atom.
%
%
%Since the exact position of the D atoms
%does not affect the computed spectra much,
%we will not consider all the possible positions
%of the D atoms, instead, we randomly locate
%the D atoms in the benzene rings.
%
In total, we consider 24 isomers
for seven parent molecules
(see Figure~\ref{fig:PAH_HD_Scheme}).

\begin{figure*}
\centering{
\includegraphics[scale=0.4,clip]{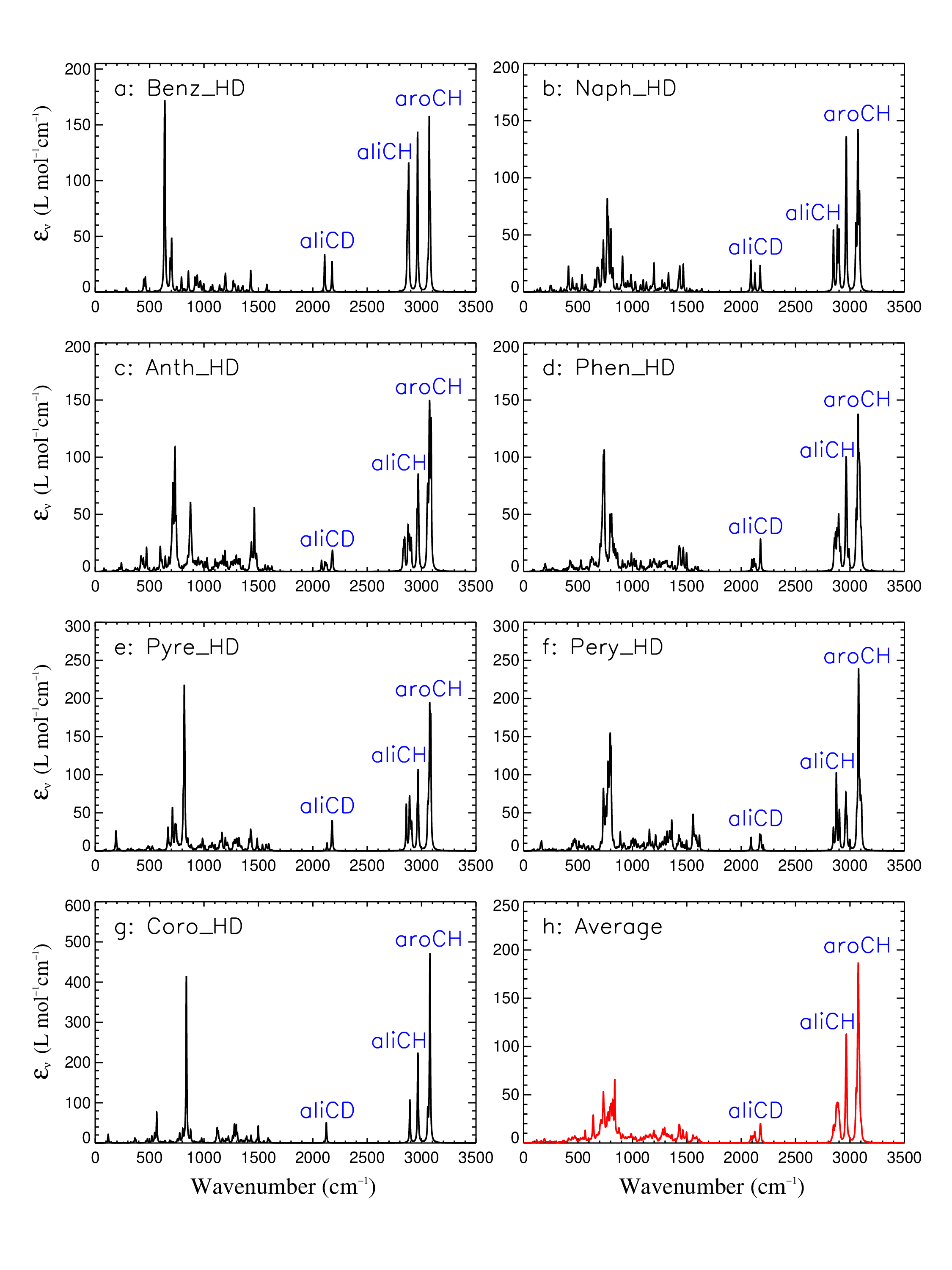}
}
\caption{\footnotesize
               \label{fig:PAH_HD_Spec}
                Same as Figure~\ref{fig:PAH_Methyl1D_Spec}
                but for ``superdeuterated'' PAHs
                in which one H atom and one D atom
                share an C atom
                (see Figure~\ref{fig:PAH_HD_Scheme}).
                }
\end{figure*}

For each parent molecule, we obtain the mean
spectrum by averaging over the isomers
(see Figures~\ref{fig:PAH_HD_Spec}a--g).
We also obtain the overall mean spectrum
by averaging over all seven parent molecules
(see Figure~\ref{fig:PAH_HD_Spec}h).
The aliphatic C--D stretches are clearly seen
at $\simali$2150$\cm^{-1}$
(i.e., $\simali$4.65$\mum$)
and exhibit multiple peaks,
indicating that the wavelengths
of the aliphatic C--D stretches
somewhat differ from one isomer to another.
%

%
%We also include three coronene derivatives, i.e.
%D atom exchange at oxidized edge sites
%or at aromatic edge sites.
%The structures of the calculated molecules
%are given in Figure~\ref{fig:PAH_HD_Scheme}.
%
%The vibrational spectra for the coronene
%derivatives are given in Figure~\ref{fig:Coro_HD_Spec}.
%The spectra show deuterium related features,
%including aliphatic and aromatic C--D stretches
%and o--D stretches.
%We note that in Coro$\_$O2D
%with two D atoms attached to
%the same C atom, the intensity of
%the aliphatic C--D stretch features is
%significantly depressed.
%

\begin{figure*}
\centering{
\includegraphics[scale=0.3,clip]{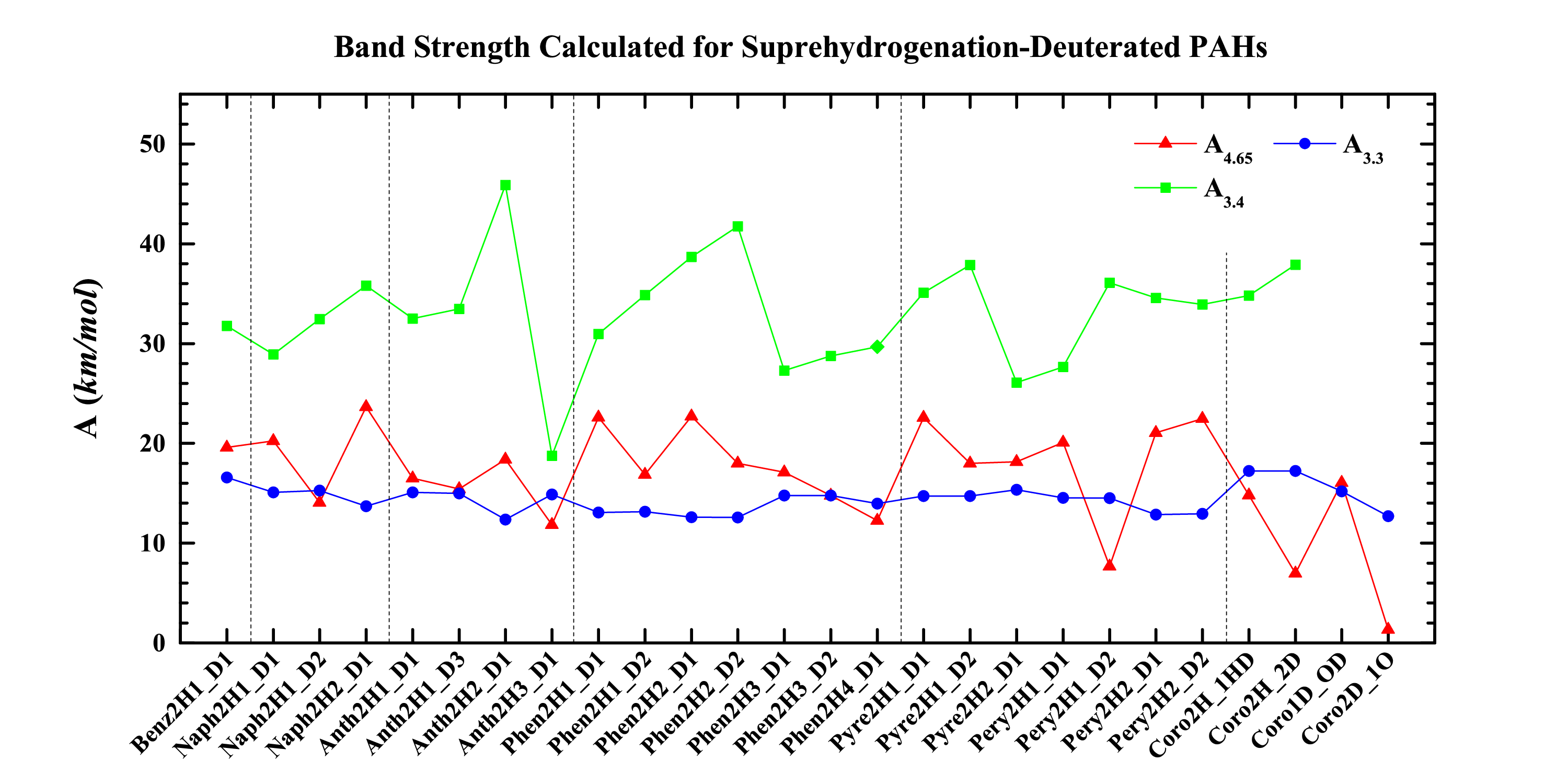}
}
\caption{\footnotesize
         \label{fig:SuperHD_Intensity}
         Same as Figure~\ref{fig:Methyl1D_Intensity}
         but for ``superdeuterated'' PAHs.
        }
\end{figure*}

The wavelengths and intensities
of the C--H and C--D stretching transitions
calculated for superdeuterated PAHs
are tabulated in Table~\ref{tab:Freq_Int_PAH_HD}
and shown in Figure~\ref{fig:SuperHD_Intensity}.
Similar to PAHs attached with a D-substituted
methyl group (PAH$\_$CH$_{2}$D),
the scatters in intensities are quite small
among isomers and parent molecules,
especially for the aromatic C--H stretch.
The aliphatic C--D stretches
of superdeuterated PAHs are more intense
and peak at longer wavelengths
than that of PAH$\_$CH$_{2}$D,
with $\langle\lambda_{\rm peak}\rangle\approx 4.66\mum$
and $\langle \Acd\rangle \approx 16.2\km\mol^{-1}$
for superdeuterated PAHs, while
$\langle\lambda_{\rm peak}\rangle\approx 4.60\mum$ and
$\langle \Acd\rangle \approx 12.9\km\mol^{-1}$
for PAH$\_$CH$_{2}$D.

\begin{figure*}
\centering{
\includegraphics[scale=0.4,clip]{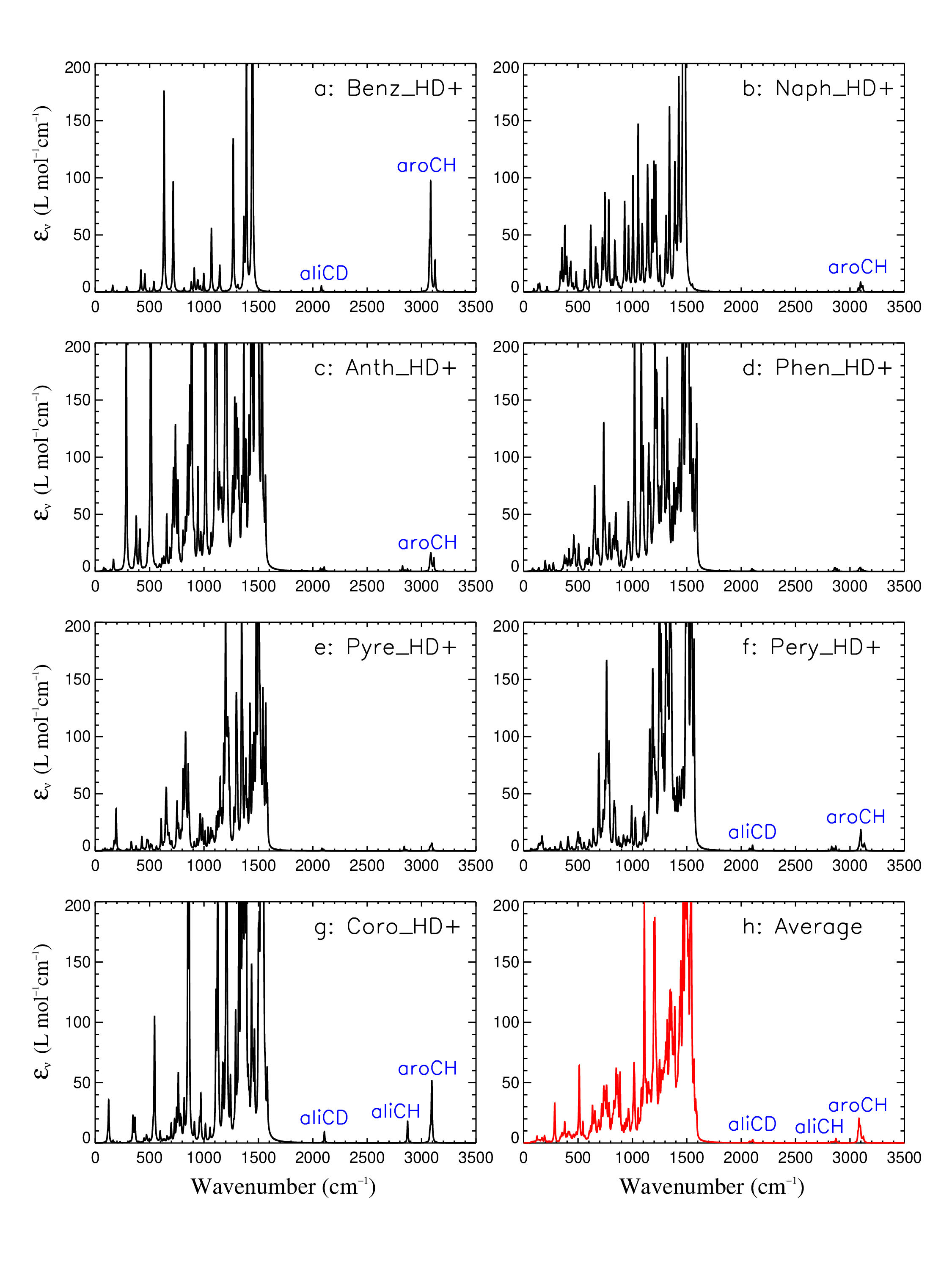}
}
\caption{\footnotesize
              \label{fig:PAH_HDPlus_spec}
              Same as Figure~\ref{fig:PAH_Methyl1D_Spec}
              but for ``superdeuterated'' PAH cations
              (see Figure~\ref{fig:PAH_HDPlus_Scheme}).
              }
\end{figure*}

\begin{figure*}
\centering{
\includegraphics[scale=0.3,clip]{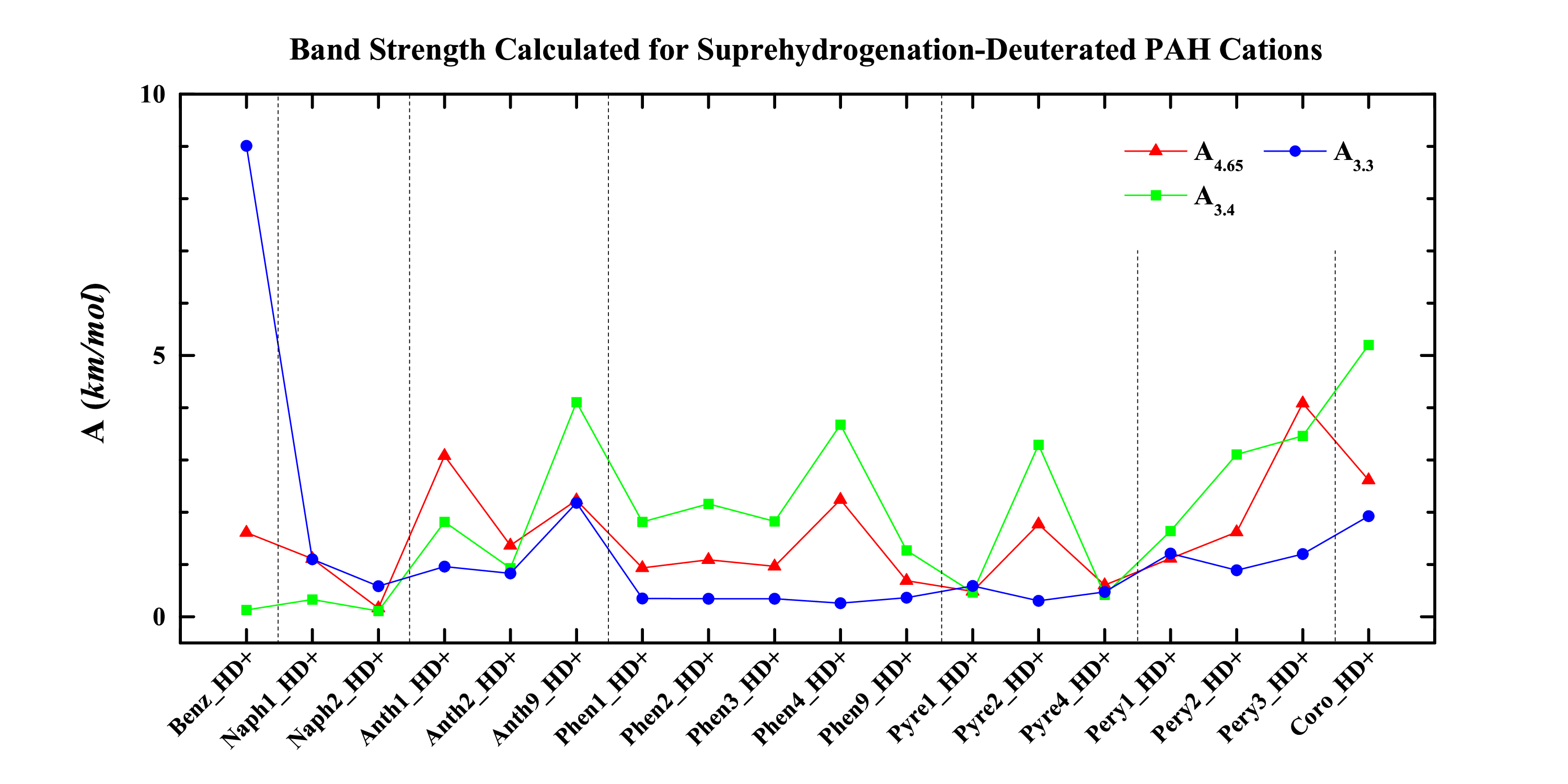}
}
\caption{\footnotesize
  \label{fig:SuperHDPlus_Intensity}
         Same as Figure~\ref{fig:Methyl1D_Intensity}
         but for ``superdeuterated'' PAH cations.
         }
\end{figure*}

To investigate the effects of ionization
on the vibrational spectra, we calculate
the spectra for ``superdeuterated'' PAH
cations (PAH$\_$HD+).
We consider the situation that only one
C atom on the edge is superhydrogenated
and attached with a pair of H and D atoms.
% a D atom squeeze in the PAH molecule and share a C atom
% on the edge with the H atom, thus forming superhydrogenation.
We consider seven parent molecules
and in total 18 isomers, with the H and D pair
attached at different positions
(see Figure~\ref{fig:PAH_HDPlus_Scheme}).
For each parent molecule,
the mean spectrum averaged over all the isomers
is given in Figures~\ref{fig:PAH_HDPlus_spec}a--g.
The overall mean spectrum averaged over
seven parent molecules is shown in
Figure~\ref{fig:PAH_HDPlus_spec}h.
It is apparent that, upon ionization,
the C--H and C--D stretches are
substantially depressed, while the C--C strecthes
at $\simali$6--9$\mum$ are appreciably enhanced.
% To show the C--H and C--D stretch features
%as much as possible, we cut the y-axis off at 200.
%Even so, only Benz$\_$HD+ and Coro$\_$HD+
%clearly show the C--H and C--D stretch features.
%
The wavelengths and intensities
of the C--H and C--D stretches
are tabulated in Table~\ref{tab:Freq_Int_PAH_HDPlus}
and shown in Figure~\ref{fig:SuperHDPlus_Intensity}.
With $\langle\lambda_{\rm peak}\rangle\approx 4.76\pm0.05\mum$
and $\langle \Acd\rangle \approx 1.64\pm0.75\km\mol^{-1}$,
the aliphatic C--D stretches of ``superdeuterated'' PAH cations
peak at significantly  longer wavelengths
and are $\simali$10 times weaker than
that of neutrals.

\subsection{PAHs with Both Aromatic and
                    Aliphatic C--D Bonds
                    \label{subsec:PAH_2D_Results}}
%
%In view of the fact that, observationally, the aliphatic
%C--D emission at $\simali$4.6--4.8$\mum$ is
%always seen together with the aromatic C--D emission
%at $\simali$4.4$\mum$ (Peeters et al.\ 2004;
%Onaka et al.\ 2014, 2022; Doney et al.\ 2016),
%we also consider PAHs with both aromatic
%and aliphatic C--D bonds,
%taking pyrene as an example.
%We consider pyrenes in which
%an D atom substitutes an H atom
%{\it and} a deuterated methyl group
%(--CH$_2$D) is also attached
%(hereafter PyreD$\_$CH$_{2}$D;
%see Figure~\ref{fig:PAH_Methyl2D_Scheme}).
%We also consider mono-deuterated pyrenes
%which are also ``superdeuterated''
%(hereafter PyreD$\_$HD;
%see Figure~\ref{fig:PAH_H2D_Scheme}).
%Again, as D is much less abundant than H,
%``superdeuterated'' pyrenes are very likely
%also superhydrogenated.
%%
%Different isomers are considered,
%with D atoms and functional groups
%attached at different positions.
%%
%
%
Considering that the aliphatic C--D bands
at $\simali$4.6--4.8$\mum$ are always seen
in emission together with the aromatic C--D bands
at $\simali$4.4$\mum$ (Peeters et al.\ 2004;
Onaka et al.\ 2014, 2022; Doney et al.\ 2016),
we also calculate the vibrational spectra of PAHs
with both aliphatic and aromatic C--D bonds,
with pyrene as an illustrative example.
We consider two kinds of aliphatic C--D units,
with the aliphatic D atom either in a methyl group
(hereafter PyreD$\_$CH$_{2}$D,
see Figure~\ref{fig:PAH_Methyl2D_Scheme})
or in a ``superdeuterated'' pair
(i.e., one D atom and one H atom share an C atom;
hereafter PyreD$\_$HD, see Figure~\ref{fig:PAH_H2D_Scheme}).
We consider 23 isomers for PyreD$\_$CH$_2$D,
with the deuterated-mehtyl group attached at
position 1, 2 or 4, and the other D atom bonding
with one of the remaining edge C atoms in pyrene.
For PyreD$\_$HD, we consider 24 isomers,
with the H+H pair and the H+D pair attached
at positions 1 and 2, 2 and 3, or 4 and 5,
while the other D atom, again, bonding with
one of the remaining edge C atoms in the rings.

\begin{figure*}
\centering{
\includegraphics[scale=0.3,clip]{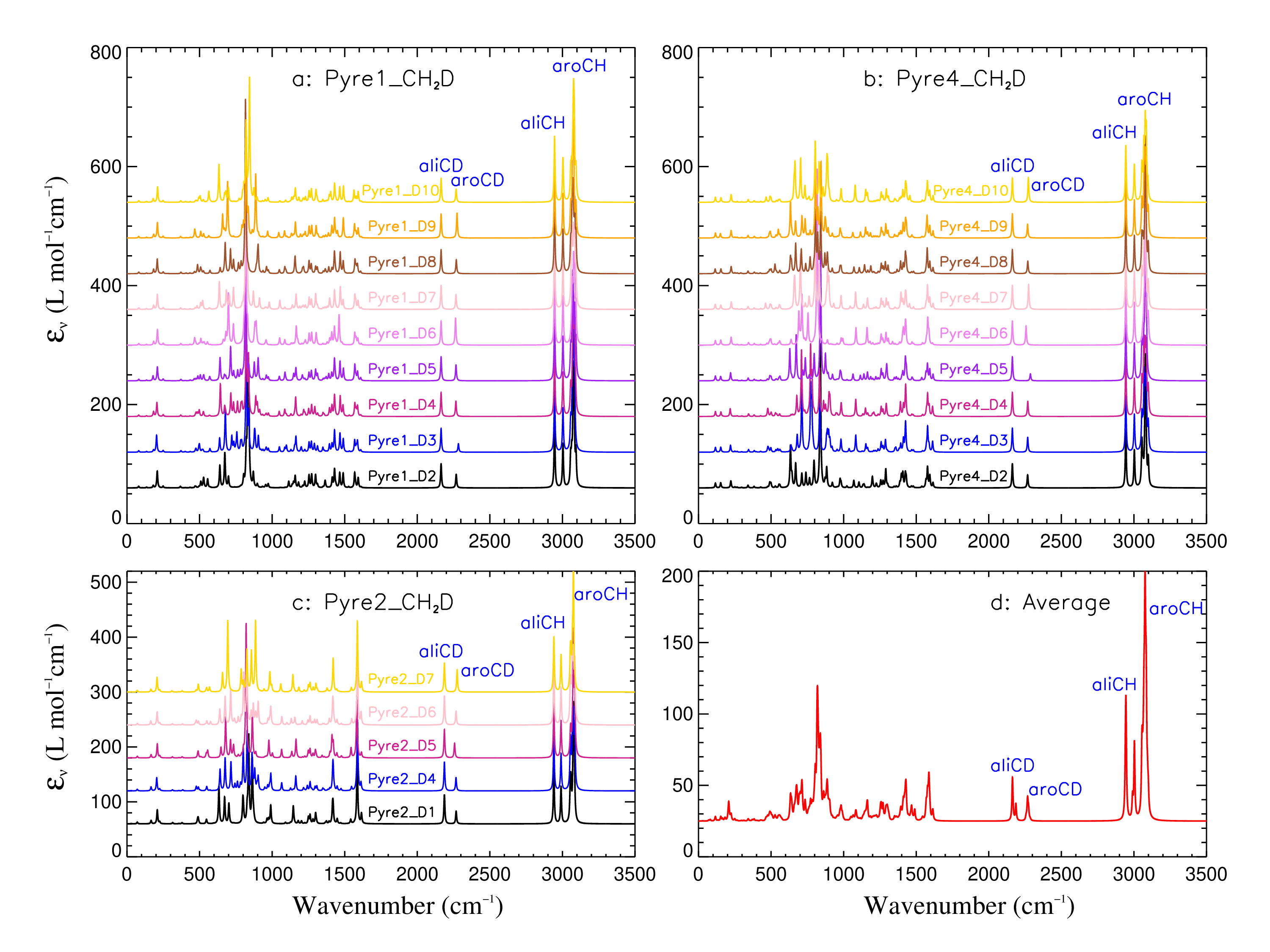}
}
\caption{\footnotesize
  \label{fig:PAH_Methyl2D_Spec}
         Same as Figure~\ref{fig:PAH_Methyl1D_Spec}
         but for pyrenes containing a methyl-deuterated
         sidegroup as well as a periphal D atom
         (see Figure~\ref{fig:PAH_Methyl2D_Scheme}).
         }
\end{figure*}

By categorizing the isomers according to
the positions of the aliphatic D atom,
we show in Figure~\ref{fig:PAH_Methyl2D_Spec}
the spectra of the isomers in each category.
We also obtain the mean spectrum
by averaging over all the somers
and show in Figure~\ref{fig:PAH_H2D_Spec}.
The aliphatic and aromatic C--D stretches
respectively at 2170$\cm^{-1}$ (i.e., 4.61$\mum$)
and 2250$\cm^{-1}$ (i.e., 4.44$\mum$)
are clearly seen in the spectra of isomers
as well as in the mean spectrum,
with the aliphatic C--D stretch
being appreciably more intense
than the aromatic C--D stretch.
It is interesting to note that the spectra
of different isomers are quite similar
(see Figures~\ref{fig:PAH_Methyl2D_Spec},\,\ref{fig:PAH_H2D_Spec}),
implying that the positions of the D atoms
do not significantly affect the spectra.

\begin{figure*}
\centering{
\includegraphics[scale=0.3,clip]{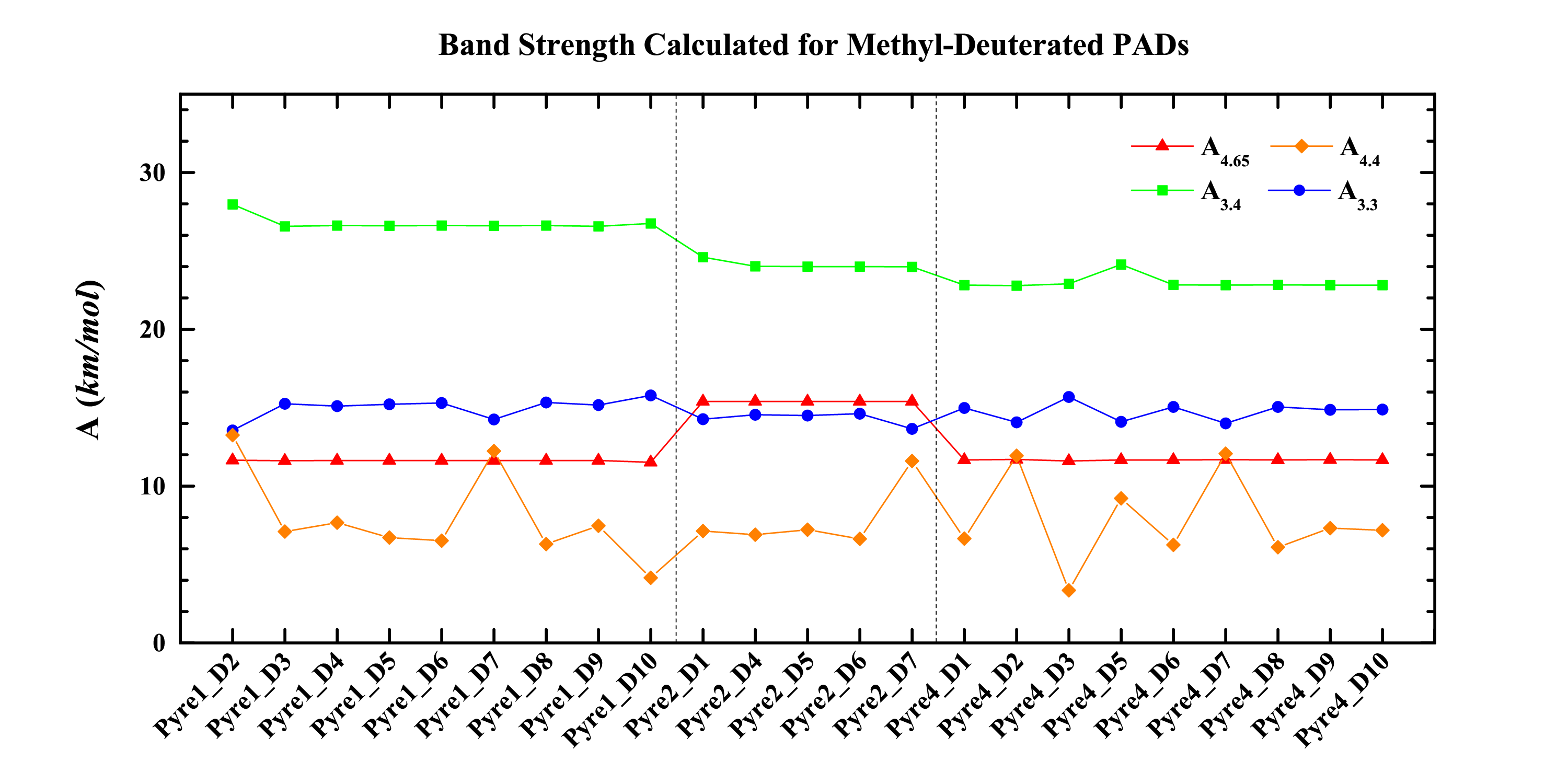}
}
\caption{\footnotesize
  \label{fig:Methyl_2D_Intensity}
         Same as Figure~\ref{fig:Methyl1D_Intensity}
         but for pyrenes containing a methyl-deuterated
         sidegroup as well as a periphal D atom.
         }
\end{figure*}

\begin{figure*}
\centering{
\includegraphics[scale=0.3,clip]{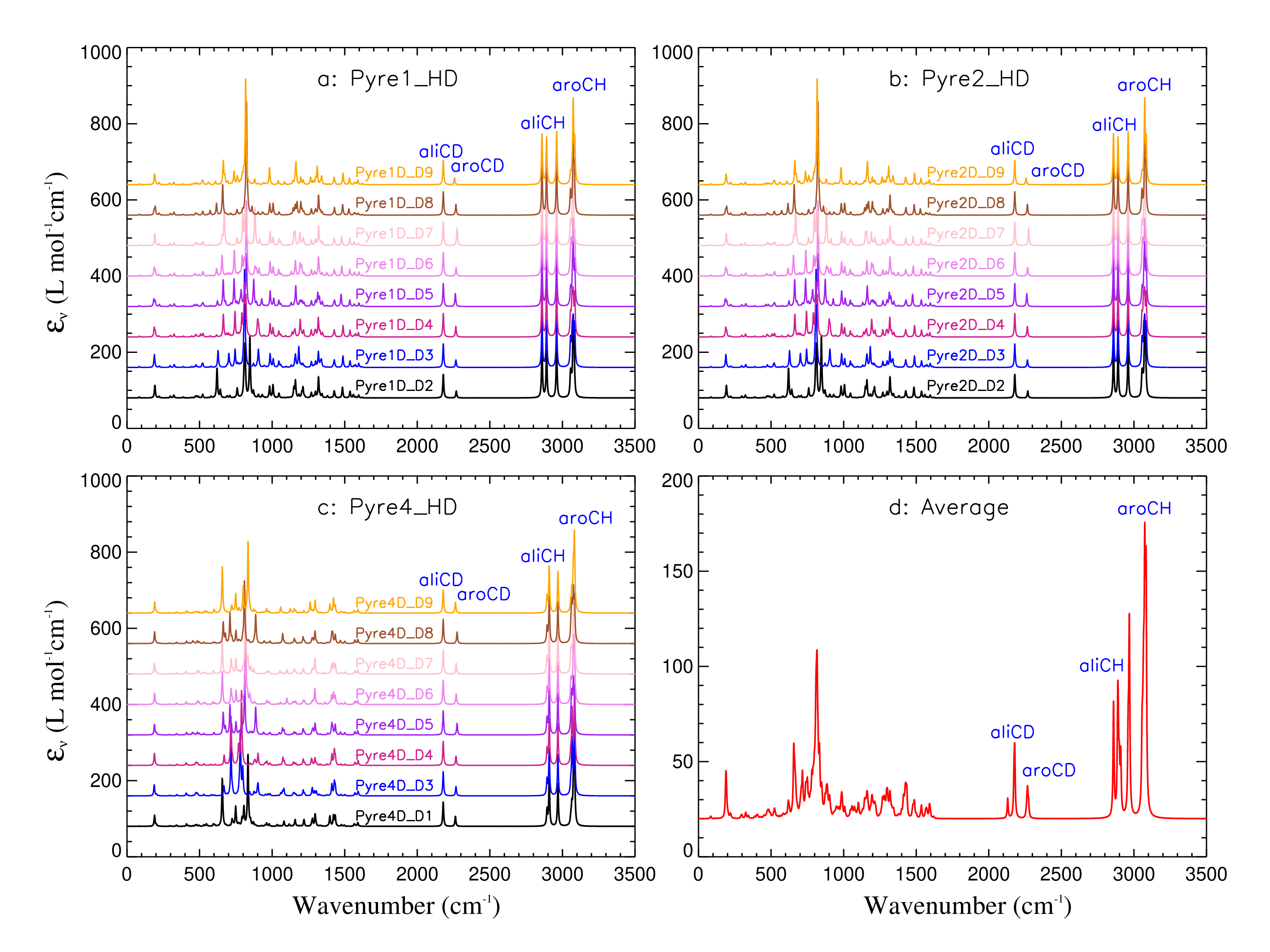}
}
\caption{\footnotesize
         \label{fig:PAH_H2D_Spec}
         Same as Figure~\ref{fig:PAH_Methyl1D_Spec}
         but for ``superdeuterated'' pyrenes
         containing an H+H pair, an H+D pair,
         as well as a periphal D atom
         (see Figure~\ref{fig:PAH_H2D_Scheme}).
         }
\end{figure*}

\begin{figure*}
\centering{
\includegraphics[scale=0.3,clip]{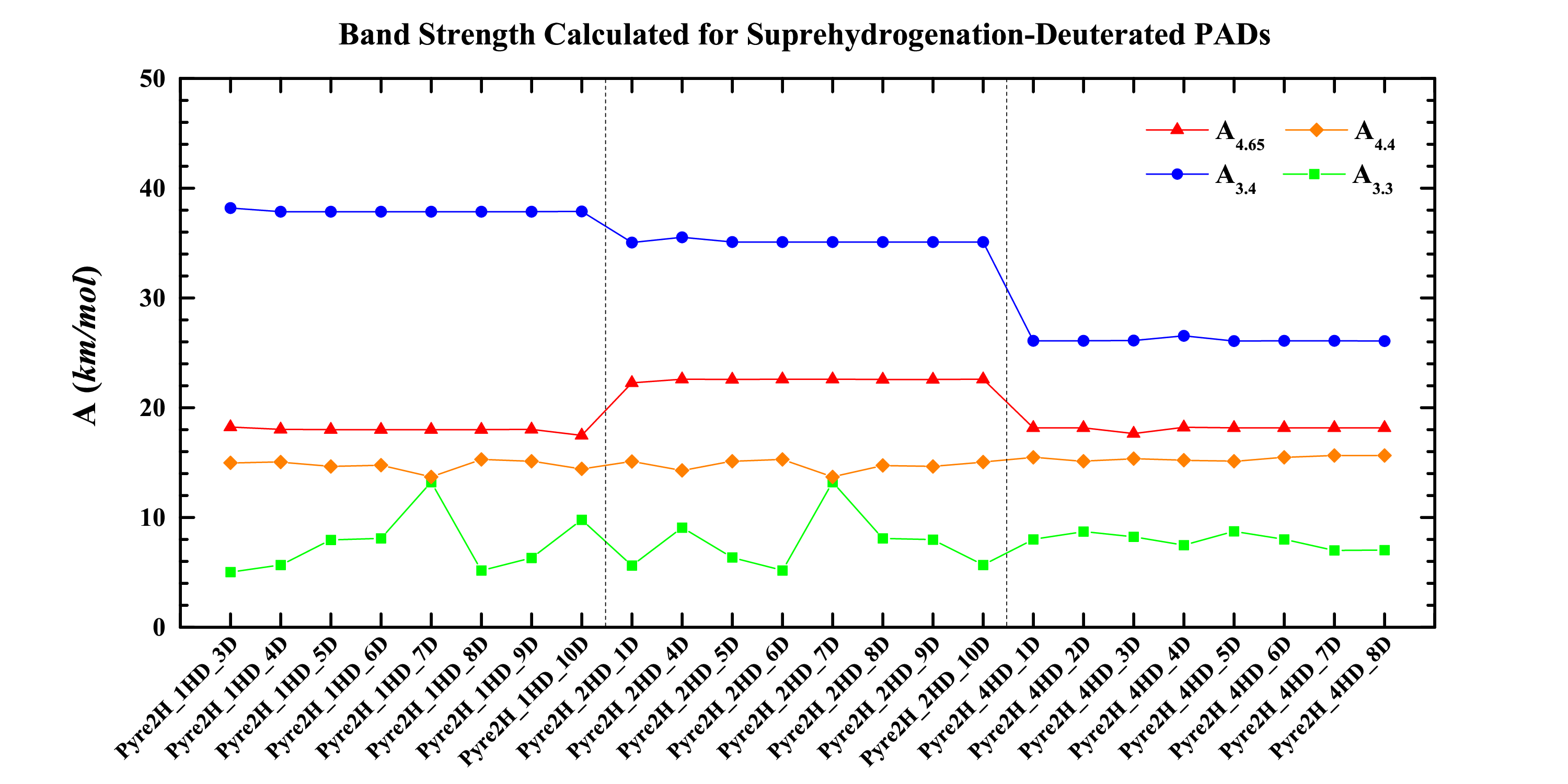}
}
\caption{\footnotesize
               \label{fig:SuperHD_2D_Intensity}
               Same as Figure~\ref{fig:Methyl1D_Intensity}
               but for ``superdeuterated'' pyrenes
               containing an H+H pair, an H+D pair,
               as well as a periphal D atom.
               }
\end{figure*}

The wavelengths and intensities of the C--H
and C--D stretches of PyreD$\_$CH$_{2}$D
and PyreD$\_$HD are respectively tabulated in
Tables~\ref{tab:Freq_Int_PAH_Methyl2D},\,\ref{tab:Freq_Int_PAH_H2D},
and the intensities are shown in
Figures~\ref{fig:Methyl_2D_Intensity},\,\ref{fig:SuperHD_2D_Intensity}.
With $\langle\lambda_{\rm peak}\rangle\approx 4.41\mum$
and $\langle \Acd\rangle \approx7.9\km\mol^{-1}$,
the wavelengths and intensities of
the aromatic C--D stretch in PyreD$\_$CH$_{2}$D
are essentially consistent with that of PyreD$\_$HD.
This is not surprising because,
according to the Born-Oppenheimer approximation,
the presence of an D-substituted methyl sidegroup
or a pair of H and D atoms sharing one C atom
would not affect the aromatic C--D stretches.
On the other hand, the intensities of the aliphatic
C--D stretches of PyreD$\_$CH$_{2}$D
($\langle \Acd\rangle \approx12.4\km\mol^{-1}$)
differ appreciably from that of PyreD$\_$HD
($\langle \Acd\rangle \approx19.5\km\mol^{-1}$).
This is understandable since they arise from
different types of aliphatic C--D units.
%
%Note here that the different groups of isomers have
%different intensities, which originates from the different
%positions of the D atoms in the methyl
%or superhydrogenation group.

\subsection{Aliphatic C--D Band Intensities}\label{sec:BandStrengths}
\begin{figure*}
\centering{
\includegraphics[scale=0.3,clip]{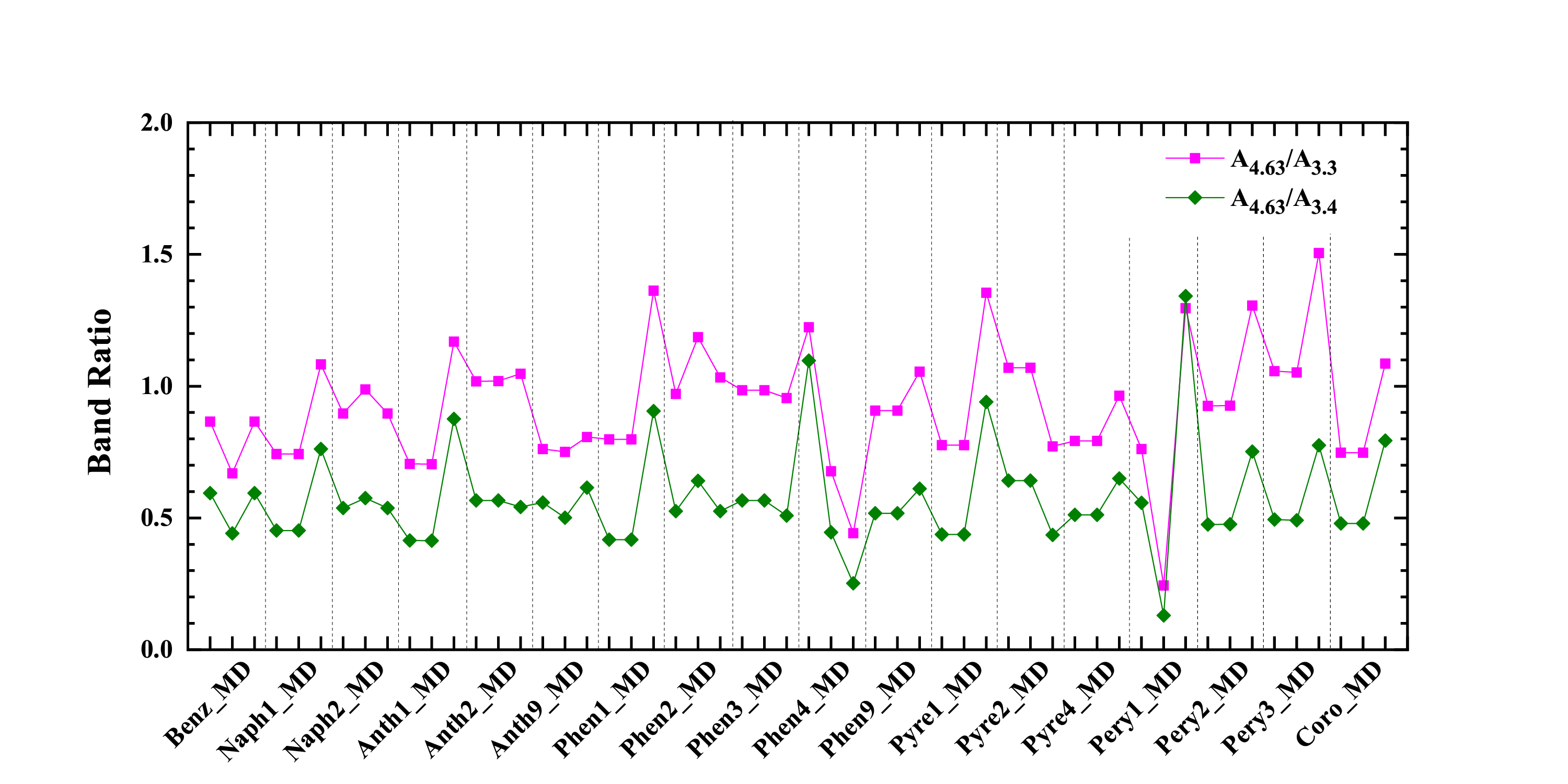}
}
\caption{\footnotesize
         \label{fig:Methyl1D_IRatio}
         Band-strength ratios of $\Acd/\Aaro$ (magenta)
         and $\Acd/\Aali$ (green)
         for methyl-deuterated PAHs
         (see Figure~\ref{fig:PAH_MethylD_Scheme}).
         }
\end{figure*}

\begin{figure*}
\centering{
\includegraphics[scale=0.3,clip]{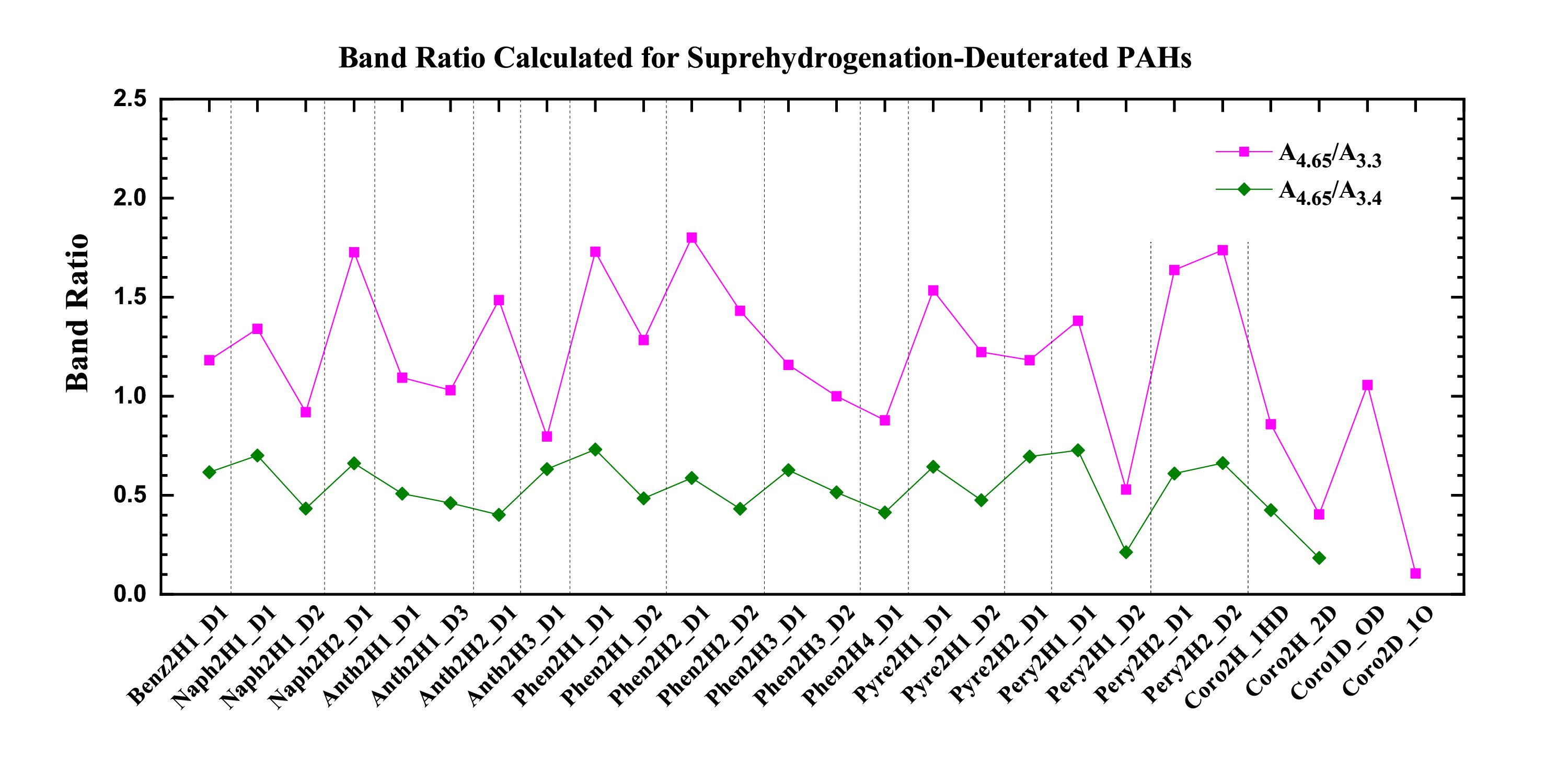}
}
\caption{\footnotesize
   \label{fig:SuperHD_IRatio}
   Band-strength ratios of $\Acd/\Aaro$ (magenta)
   and $\Acd/\Aali$ (green)
   for ``superdeuterated'' PAHs
   in which one H atom and one D atom
   share an C atom
   (see Figure~\ref{fig:PAH_HD_Scheme}).
         }
\end{figure*}

\begin{figure*}
\centering{
\includegraphics[scale=0.3,clip]{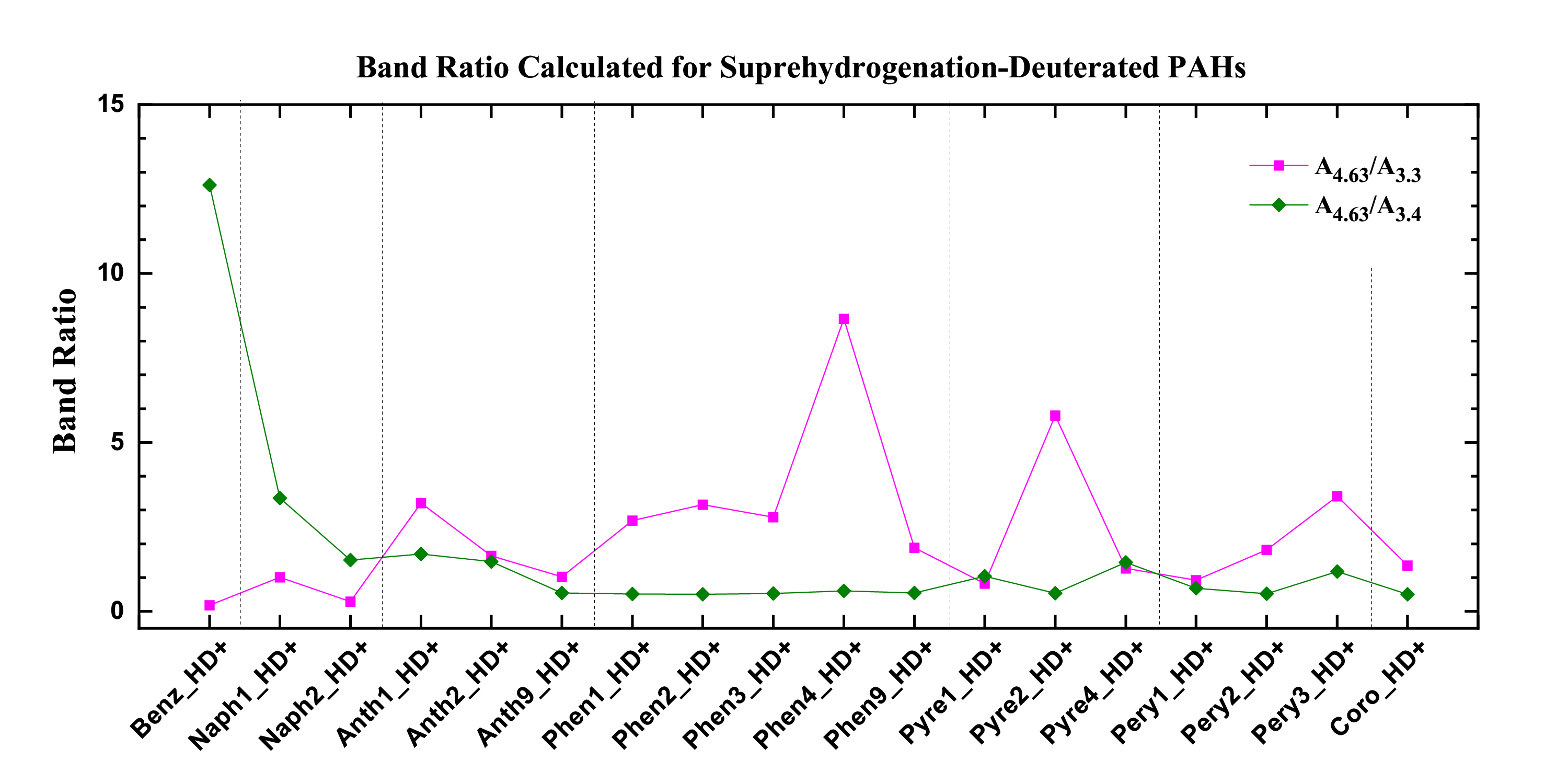}
}
\caption{\footnotesize
       \label{fig:SuperHDPlus_IRatio}
       Band-strength ratios of $\Acd/\Aaro$ (magenta)
       and $\Acd/\Aali$ (green)
       for ``superdeuterated'' PAH cations
       (see Figure~\ref{fig:PAH_HDPlus_Scheme}).
       }
\end{figure*}

\begin{figure*}
\centering{
\includegraphics[scale=0.3,clip]{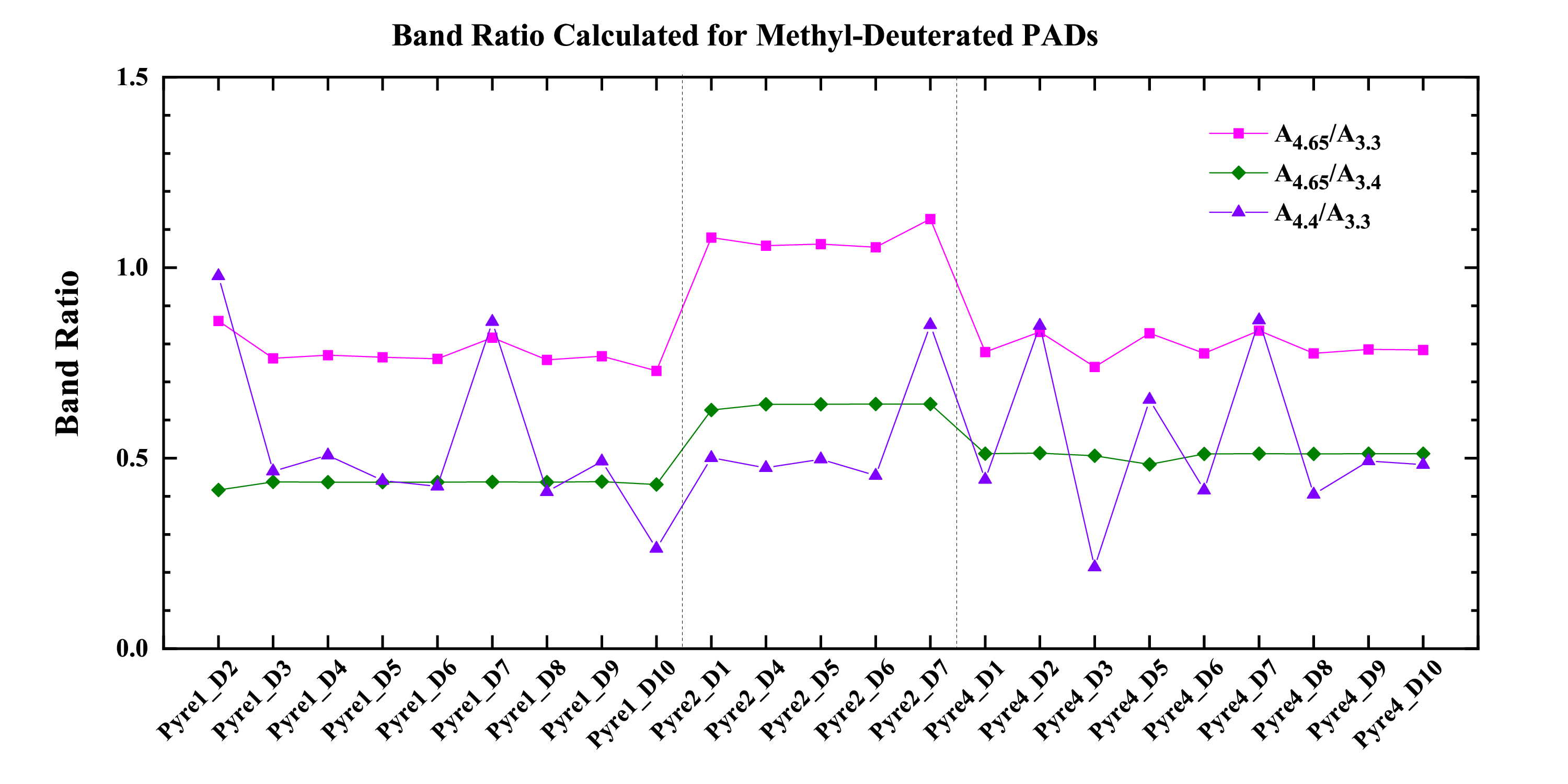}
}
\caption{\footnotesize
   \label{fig:Methyl2D_IRatio}
   Band-strength ratios of $\Acd/\Aaro$ (magenta)
   $\Acd/\Aali$ (green), and $\ACD/\Aaro$ (blue)
   for pyrenes containing a methyl-deuterated
   sidegroup as well as a periphal D atom
   (see Figure~\ref{fig:PAH_Methyl2D_Scheme}).
         }
\end{figure*}

\begin{figure*}
\centering{
\includegraphics[scale=0.3,clip]{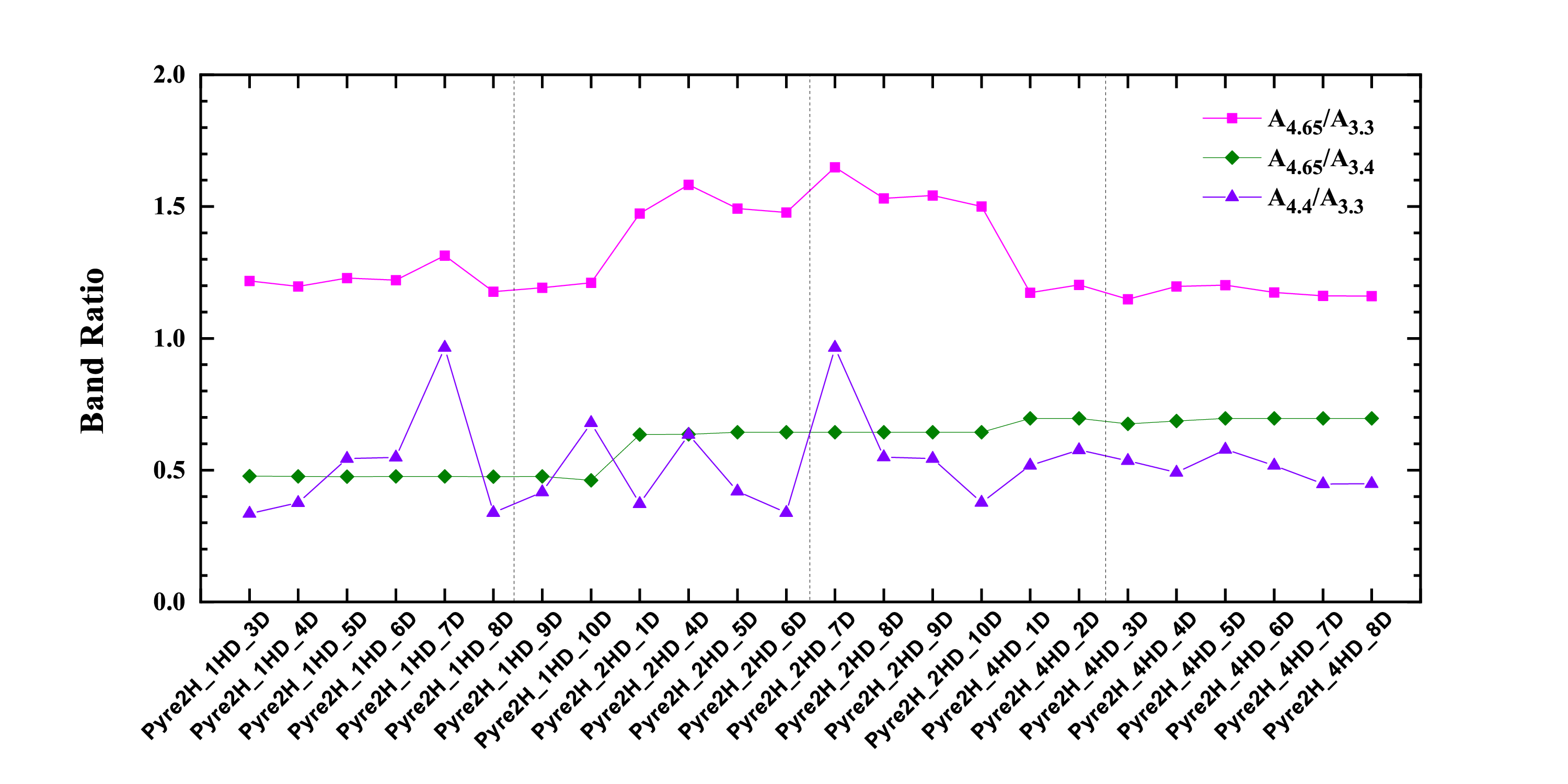}
}
\caption{\footnotesize
   \label{fig:SuperHD_2D_IRatio}
   Band-strength ratios of $\Acd/\Aaro$ (magenta)
   $\Acd/\Aali$ (green), and $\ACD/\Aaro$ (blue)
   for ``superdeuterated'' pyrenes
         containing an H+H pair, an H+D pair,
         as well as a periphal D atom
         (see Figure~\ref{fig:PAH_H2D_Scheme}).
         }
\end{figure*}

In Figures~\ref{fig:Methyl1D_IRatio}--\ref{fig:SuperHD_2D_IRatio},
we show $\Aratio$---the band-intensity ratio of
the 4.65$\mum$ aliphatic C--D stretch to that of
the 3.3$\mum$ aromatic C--H stretch---calculated
for each set of our target molecules.
Generally speaking, $\Aratio$ does not vary much
from one molecule to another in each set.
We obtain $\langle\Aratio\rangle\approx0.90\pm0.07$
for PAHs attached with a D-substituted methyl sidegroup
(PAH$\_$CH$_2$D), and
$\langle\Aratio\rangle\approx1.10\pm0.32$
for ``superdeuterated'' PAHs (PAH$\_$HD)
in which an extra D atom shares an C atom
with an H atom.
For pyrenes attached with a D-substituted methyl
sidegroup (PyreD$\_$CH$_2$D), we obtain
$\langle\Aratio\rangle\approx0.85$
with a standard deviation of $\simali$0.12;
for ``superdeuterated'' pyrenes (PyreD$\_$HD),
we obtain $\langle\Aratio\rangle\approx1.31$
with a standard deviation of $\simali$0.17.

In Table~\ref{tab:MeanBandStrengths} we summarize
the {\it mean} band strengths of the aliphatic C--D
and aromatic C--H stretches, obtained by averaging
over all the species considered here.
The recommended band intensities are
$\langle\Acd\rangle\approx15.3\pm3.3\km\mol^{-1}$,
$\langle\Aaro\rangle\approx14.8\pm0.33\km\mol^{-1}$,
and $\langle\Acd/\Aaro\rangle\approx1.04\pm0.21$
for deuterated neutral PAHs, and
$\langle\Acd\rangle\approx1.64\pm0.75\km\mol^{-1}$,
$\langle\Aaro\rangle\approx2.14^{+3.08}_{-2.14}\km\mol^{-1}$,
and $\langle\Acd/\Aaro\rangle\approx1.67\pm1.00$
for deuterated PAH cations.\footnote{%
  The large scatter in $\langle\Aaro\rangle$
    for PAH cations mainly arises from
    the large $\Aaro$ intensity of Benz$\_$HD+
    (see Table~\ref{tab:Freq_Int_PAH_HDPlus}),
    which is substantially larger than that of the other
    molecules. If we exclude Benz$\_$HD+, we obtain
    $\langle\Aaro\rangle\approx1.00\pm0.59\km\mol^{-1}$
    and $\langle\Acd/\Aaro\rangle\approx1.92\pm0.94$.
    However, this does not affect the degree of deuteration
    of PAHs derived below in \S\ref{sec:dpah}
    since the C--D and C--H stretches are dominated
    by neutral PAHs
    }
Since the intensities for the C--H or C--D stretches
are significantly depressed for PAH cations,
when estimating the D enrichment
from astronomical observations,
we recommend to adopt $\Aratio\approx1.04\pm0.21$,
the band-intensity ratio of deuterated neutral PAHs.

\section{Astrophysical Applications}\label{sec:astro}
\subsection{Aliphatic C--D Bands:
                    Astronomical Observations}\label{sec:Obs}
Observational efforts have been made
with the {\it Short Wavelength Spectrometer} (SWS)
on board the {\it Infrared Space Observatory} (ISO)
and the {\it Infrared Camera} (IRC) on board AKARI
to search for signals of deuterated PAHs
through their vibrational C--D bands
(Verstraete et al.\ 1996; Peeters et al.\ 2004;
Onaka et al.\ 2014, 2022; Doney et al.\ 2016).
So far, the detetcion of the aliphatic C--D bands
has been reported in eight sources
(Peeters et al.\ 2004, Onaka et al.\ 2014, Doney et al.\ 2016).
We compile all the available observational data
for the 3.3$\mum$ aromatic C--H stretch,
the 3.4$\mum$ aliphatic C--H stretch,
the 4.4$\mum$ aromatic C--D stretch, and
the ``4.65''$\mum$ aliphatic C--D stretch.\footnote{%
  The aliphatic C--D stretch of PAHs,
  consisting of asymmetric and symmetric modes,
  actually spans a range of wavelengths,
  from $\simali$4.6$\mum$ to $\simali$4.8$\mum$.
  In this work, by ``4.65$\mum$'' we actually
  mean the aliphatic C--D stretch
  over the wavelength range of
  $\simali$4.6--4.8$\mum$.
   }
The observational emission intensities of these
stretches ($I_{3.3}$, $I_{3.4}$, $I_{4.4}$, and $I_{4.65}$)
and their ratios with respect to the 3.3$\mum$
aromatic C--H stretch,
$\left(I_{4.65}/I_{3.3}\right)_{\rm obs}$,
$\left(I_{4.4}/I_{3.3}\right)_{\rm obs}$ and
$\left(I_{3.4}/I_{3.3}\right)_{\rm obs}$
are summarized in Table~\ref{tab:I465I33_obs}.

\begin{figure*}
\centering{
\includegraphics[scale=0.6,clip]{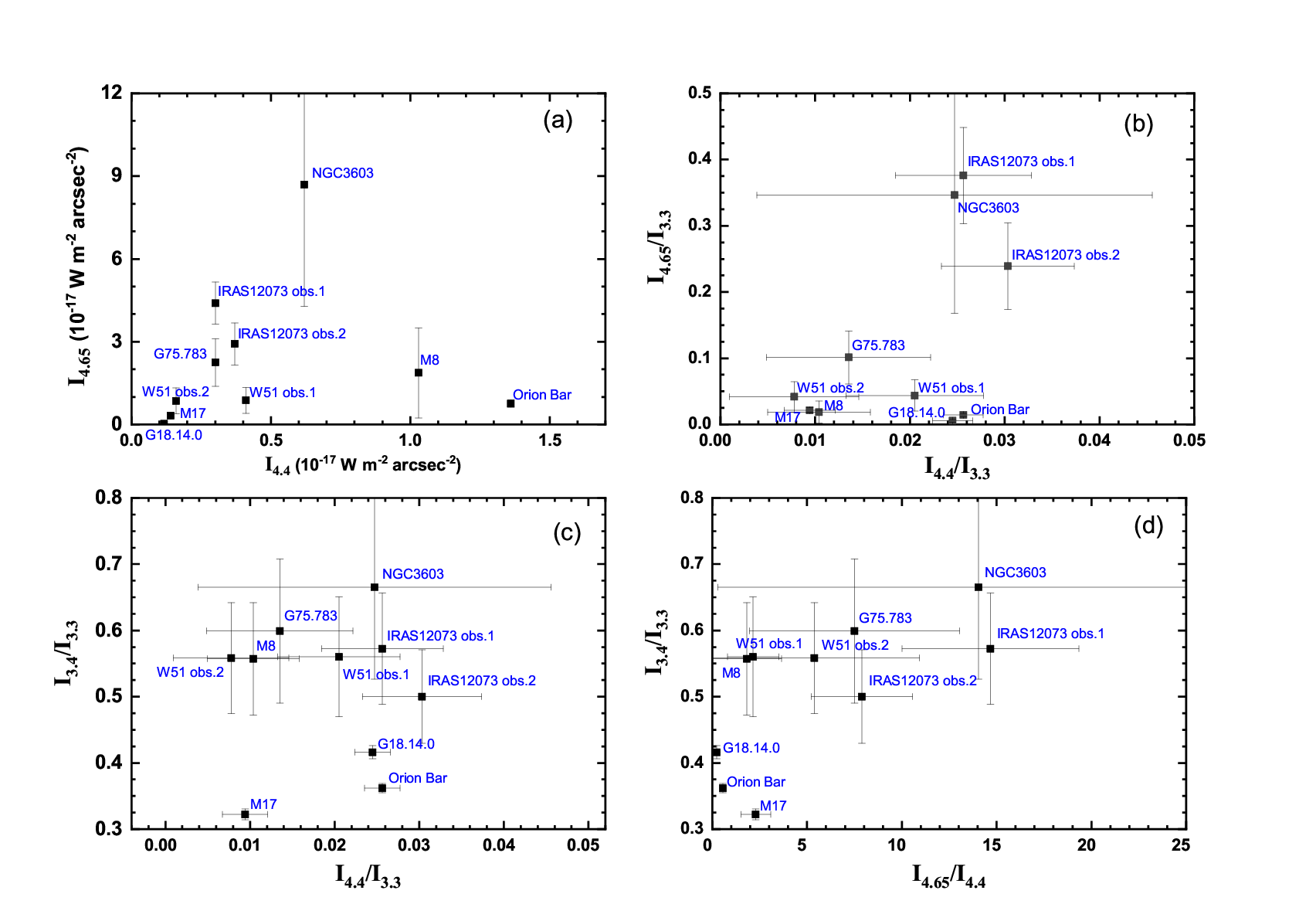}
}
\caption{\footnotesize
    \label{fig:CD_obs_data}
    Upper left (a): Correlation of the observed intensities
    of the 4.65$\mum$ aliphatic C--D band
    ($I_{4.65}$) with that of the 4.4$\mum$ aromatic
    C--D band ($I_{4.4}$).
    Upper right (b): Correlation of $I_{4.65}/I_{3.3}$
    with $I_{4.4}/I_{3.3}$.
    Bottom left (c): Correlation of $I_{3.4}/I_{3.3}$
    with $I_{4.4}/I_{3.3}$.
    Bottom right (d): Correlation of $I_{3.4}/I_{3.3}$
    with $I_{4.65}/I_{4.4}$.
         }
\end{figure*}

We explore whether these stretching bands
are somewhat related to each other.
Figure~\ref{fig:CD_obs_data}a compares
the observed emission intensities of
the aliphatic C--D stretch ($I_{4.65}$) with
that of the aromatic C--D stretch ($I_{4.4}$).
No correlation is found between $I_{4.65}$ and $I_{4.4}$.
To eliminate their common correlation with
the starlight radiation field (i.e., the observed
emission intensities of all these bands are
proportional to the starlight intensity),
we normalize $I_{4.65}$ and $I_{4.4}$
by $I_{3.3}$ and then compare $I_{4.65}/I_{3.3}$
with  $I_{4.4}/I_{3.3}$.
As demonstrated in Figure~\ref{fig:CD_obs_data}b,
$I_{4.65}/I_{3.3}$ and $I_{4.4}/I_{3.3}$ are not correlated.
This indicates that the deuteration of the aliphatic
C atoms may not occur together with the deuteration
of the aromatic C atoms (i.e., the aliphatic and aromatic
C atoms get deuterated independently).
Also, to examine the degree of deuteration of
the aliphatic C atoms and that of the aromatic
C atoms, we compare $I_{4.65}/I_{3.4}$ with $I_{4.4}/I_{3.3}$
and find no correlation either
(see Figure~\ref{fig:CD_obs_data}c).
This also implies that the deuterations
of the aliphatic and aromatic C atoms
occur independently.
Finally, we compare $I_{4.65}/I_{4.4}$
with $I_{3.4}/I_{3.3}$. This is to investigate
if the aliphaticity of the C--D bond
is related to that of the C--H bond.
As shown in Figure~\ref{fig:CD_obs_data}d,
they do not correlate.

\subsection{Deuterium Depletion onto PAHs}\label{sec:dpah}
We now evaluate from the observed emission
intensities of the 4.65$\mum$ aliphatic C--D
stretch the number of D atoms
(relative to H) that could have been locked up
in the aliphatic C--D units of PAHs.
%
%Let $\faliCD$ be the degree of deuteration
%(i.e., the total amount of D relative to H)
%of the aliphatic C--D unis of PAHs.
For a PAH molecule containing $\NH$ aromatic H atoms,
and $\ND$ aliphatic D atoms, we define
the degree of deuteration of the molecule
as $\faliCD\equiv\ND/\left(\NH+\ND\right)$,
the fraction of the total (H and D) atoms in the form of D.
The {\it observed} ratio of the power emitted
from the 4.65$\mum$ aliphatic C--D band
to that from the 3.3$\mum$ aromatic C--H band
is related to the degree of deuteration through
\begin{equation}\label{eq:Iratio}
\left(\frac{I_{4.65}}{I_{3.3}}\right)_{\rm obs}
\approx \left(\frac{A_{4.65}}{A_{3.3}}\right)
\times\left(\frac{\NCD}{\NCH}\right)
\times\left(\frac{B_{4.65}}{B_{3.3}}\right) ~~,
\end{equation}
where $\NCH$ and $\NCD$ are respectively
the number of C--H and C--D bonds in
a deuterated PAH molecule,
$B_\lambda(T)$ is the Planck function
at wavelength $\lambda$ and temperature $T$,
and $A_{3.3}$ and $A_{4.65}$
are the intrinsic band strengths
of the aromatic C--H and aliphatic C--D stretches
(on a per C--H or C--D bond basis).
Since the observed C--H and C--D stretching
emission bands arise predominantly from
neutral PAHs, we will adopt
$\Acd/\ACH\approx1.04\pm0.21$
calculated for neutral PAHs
containing aliphatic C--D units
(see \S\ref{sec:BandStrengths}
and Table~\ref{tab:MeanBandStrengths}).

Considering the relatively small abundance of D atoms
in space, it is unlikely that an aliphatic C atom
would be doubly deuterated
(i.e., two D atoms share one C atom).
Therefore, it is reasonable to assume
$\NCD=\ND$ and $\NCH=\NH$.
%\footnote{%
%    We note that interstellar PAHs could be superhydrogenated
%    (Bernstein et al.\ 1996, Sandford et al.\ 2013, Yang et al.\ 2020)
%    or attached with aliphatic functional groups
%    (Geballe et al.\ 1985, Sandford 1991,
%     Kwok \& Zhang 2011, Yang et al.\ 2017)
%    which could convert the aromatic C atoms to aliphatic.
%    However, we have shown that the degree of
%    superhydrogenation (Yang et al.\ 2020)
%    and the aliphatic content of PAHs are minor
%    (Li \& Draine 2012, Yang et al.\ 2013, 2017).
%    Nevertheless, the aliphatic C--D stretches
%    at 4.65 and 4.75$\mum$ appear to be present
%    in the ISO/SWS and AKARI/IRC spectra
%    of several interstellar objects
%    (Peeters et al.\ 2004, Onaka et al.\ 2014,
%    Doney et al.\ 2016),
%    indicating that some D atoms may be attached
%    to aliphatic C atoms.
%    }
%
The degree of deuteration of the molecule becomes
\begin{equation}
\label{eq:NDNH}
 \faliCD \approx
\left\{
1\,+\,\left(\frac{I_{3.3}}{I_{4.65}}\right)_{\rm obs}
\times\left(\frac{\Acd}{\Aaro}\right)
\times \left(\frac{B_{4.65}}{B_{3.3}}\right)
\right\}^{-1} ~~.
\end{equation}
The 3.3$\mum$ C--H stretch
and 4.65$\mum$ C--D stretch
are most effectively emitted respectively
by PAHs of vibrational temperatures of
$\simali$730$\K$ and $\simali$520$\K$
(see Foonote No.\,10 in Yang et al.\ 2020),
stochastically heated by individual
UV photons (Draine \& Li 2001).
%\footnote{%
%  Let $C_{\rm abs}(\lambda)$ be the dust
%  absorption cross section at wavelength $\lambda$
%  and $j_\lambda$ be the dust IR emissivity.
%  For $C_{\rm abs}(\lambda)\propto \lambda^{-\beta}$,
%  $\lambda j_\lambda$ peaks at
%  $\lambda_p\approx\left(h c/k T\right)/\left(4+\beta\right)$,
%  where $h$ is the Planck constant,
%  $c$ is the speed of light, and $k$ is
%  the Boltzmann constant (see Li 2009).
%  With $\beta\approx2$ for PAH-like molecules,
%  we have $T\approx728\K$ for $\lambda_p\sim3.3\mum$
%  and $T\approx516\K$ for $\lambda_p\sim4.65\mum$.
%  }
%%
For $520\simlt T\simlt 730\K$,
we obtain $B_{3.3}/B_{4.65}\approx0.72\pm0.14$.
By relaxing the temperature range to
$400\simlt T\simlt 900\K$,
we obtain $B_{3.3}/B_{4.65}\approx0.78\pm0.32$.
%\footnote{%
Even if we extend the temperature range to
$300\simlt T\simlt 1000\K$,
a similar ratio of
$B_{3.3}/B_{4.65}\approx0.78\pm0.44$
is still obtained.
%   }
%
In the following, we will adopt
$B_{3.3}/B_{4.65}\approx0.78\pm0.32$
that is suitable for $400\simlt T\simlt 900\K$.

With $B_{3.3}/B_{4.65}\approx0.78\pm0.32$
and $A_{4.65}/A_{3.3}\approx1.04\pm0.21$
(see \S\ref{sec:BandStrengths}),
for each of the eight sources
we obtain $\faliCD$ from $\Iratioobs$.
These sources also exhibit an aromatic
C--D stretch at $\simali$4.4$\mum$ as well
as an aliphatic C--H stretch at 3.4$\mum$.
Following Yang et al.\ (2020),
for each source we derive $\faroCD$,
the deuteration degree of PAHs
based on the aromatic C--D stretch.
Similarly, following Yang et al.\ (2013),
for each source we also derive $\faliCH$,
the aliphatic fraction of PAHs
based on the aliphatic C--H stretch.
The results are tabulated
in Table~\ref{tab:I465I33_obs}.

It is gratifying that in some sources
(e.g., NGC~3603, IRAS~12073, M17b)
$\faliCD$ can be as much as $\simali$20\%,
implying that PAHs could be a plausible sink
of the missing D atoms.
In the ISM, PAHs require a total C/H abundance
of $\cpah\approx60\ppm$ (Li \& Draine 2001),
 where $\cpah$ refers to the amount
of C locked up in PAHs
(relative to the total H in the ISM).
For small PAHs like coronene, the ratio of
the number of C atoms to the number of H atoms
is $\NC/\NH\approx2$.
Therefore, PAHs may contain
an amount of $\simali$30$\ppm$ of H
(relative to the total H in the ISM).
A combination of $\faliCD$ and $\faroCD$
indicates that in some sources up to $\simali$28\%
of the H atoms in PAHs could be deuterated
(see Table~\ref{tab:I465I33_obs}),
i.e., $\dpah\approx8.4\ppm$ --  the amount
of D locked up in PAHs
(relative to the total H in the ISM).
This exceeds the median of the missing D/H
($\simali$7$\ppm$; see \S\ref{sec:intro}).
We note that aliphatic C--D is often created
from the deuteration of methyl-group or
superdeuteration. This naturally accompanies
with at least one aliphatic C--H bond.
Therefore, one would expect a higher $\faliCH$
than $\faliCD$. Indeed, Table~\ref{tab:I465I33_obs}
shows that in some sources $\faliCH$ is twice
as much as $\faliCD$.

Finally, we should also note that the above
estimation of $\approx$\,8.4$\ppm$ of D
depleted in PAHs (relative to the total H
in the ISM) is rather generous.
This is based on the assumption
that both small ($\NC\simlt40$)
and large PAHs ($\NC\simgt100$)
are deuterated. For large PAHs,
deuteration may occur in D-enriched
ice mantles (see \S\ref{sec:routes}).
Even if they are deuterated,
due to their large heat capacities,
large PAHs are not excited sufficiently
to temperatures high enough to emit
at the C--H and C--D stretching bands.

In addition to large PAHs, hydrogenated
amorphous carbon (HAC) solids could also
deplete some of the D atoms missing from
the gas phase. Furton et al.\ (1999) found
that the interstellar 3.4$\mum$ aliphatic
C--H {\it absorption} band is most consistent
with $\simali$80$\ppm$ of C
(relative to the total H in the ISM)
with an H/C ratio of $\simali$0.5.
Compared with PAHs, HAC would provide
a larger number of potential sites for D-depletion.
It would be interesting to see if the lines of sight
exhibiting strong 3.4$\mum$ aliphatic C--H
absorption also show detectable aliphatic
C--D absorption at $\simali$4.65$\mum$.

In the local diffuse ISM,
the optical depth of the 3.4$\mum$
aliphatic C--H {\it absorption} relative to
the visual extinction ($A_V$) is
$\tauch/A_V \approx 1/274$
(Sandford et al.\ 1991,
Pendleton et al.\ 1994;
see Gao et al.\ 2010 for a summary).
Toward the Galactic center,
with $\tauch/A_V \approx 1/146$
(McFadzean et al.\ 1989,
Pendleton et al.\ 1994;
also see Gao et al.\ 2010 for a summary),
the 3.4$\mum$ absorption band
is stronger (on a per magnitude
of visual extinction basis)
by a factor of $\simali$1.9.
If we take the band-intensity ratio
of the 4.65$\mum$ aliphatic C--D
to the 3.4$\mum$ aliphatic C--H
to be $A_{4.65}/A_{3.4}\approx0.38$
as measured for deuterated quenched
carbonaceous composite materials
(Mori et al.\ 2022),
and assume that HAC contains
$\simali$40$\ppm$ of H
and $\simali$7$\ppm$ of D
(relative to the total H in the ISM),
we expect the optical depth of
the 4.65$\mum$ aliphatic C--D
absorption relative to
the visual extinction to be
$\taucd/A_V \approx 1/4150$
for the local ISM
and $\taucd/A_V \approx 1/2200$
for the Galactic center.
Such a weak band is beyond
the detection capability of ISO.
JWST's unprecedented sensitivity
will provide a unique opportunity
to detect or place an upper limit on
the 4.65$\mum$ aliphatic
C--D {\it absorption} band
in heavily obscured lines of sight.

%

% Besides, a $\faliCD$ of $\sim$ 20\%
%implies a D-enrichment in PAHs
%by a factor of $\Dpah/\Dism\approx10^4$
%by adopting interstellar abundance of
%$\Dism\approx2\times10^{-5}$.

%In PDRs, the H atoms are abundant
%and the radiation field is weak.
%The D atoms could possibly incorporated
%with superhydrogenated PAHs
%and form PAHs with deuterated
%superhydrogenation group.

%To form methyl-deuterated PAHs,
%the C--H bond in the methyl must
%be broken and then D atoms come in.
%However, ...
%Another possibility is to form
%in the molecular clouds where
%there are abundant HDO.

\subsection{Possible Routes for Aliphatically
                    Deuterating PAHs}\label{sec:routes}
A significant fraction of the UV radiation
emitted by massive stars impinges on
the molecular gas associated with star formation.
This results in a PDR bounded on one side
by an ionization front, and on the other side
by cold molecular gas which has not yet been
appreciably affected by UV radiation.

\begin{figure*}
\centering{
\includegraphics[scale=0.5,clip]{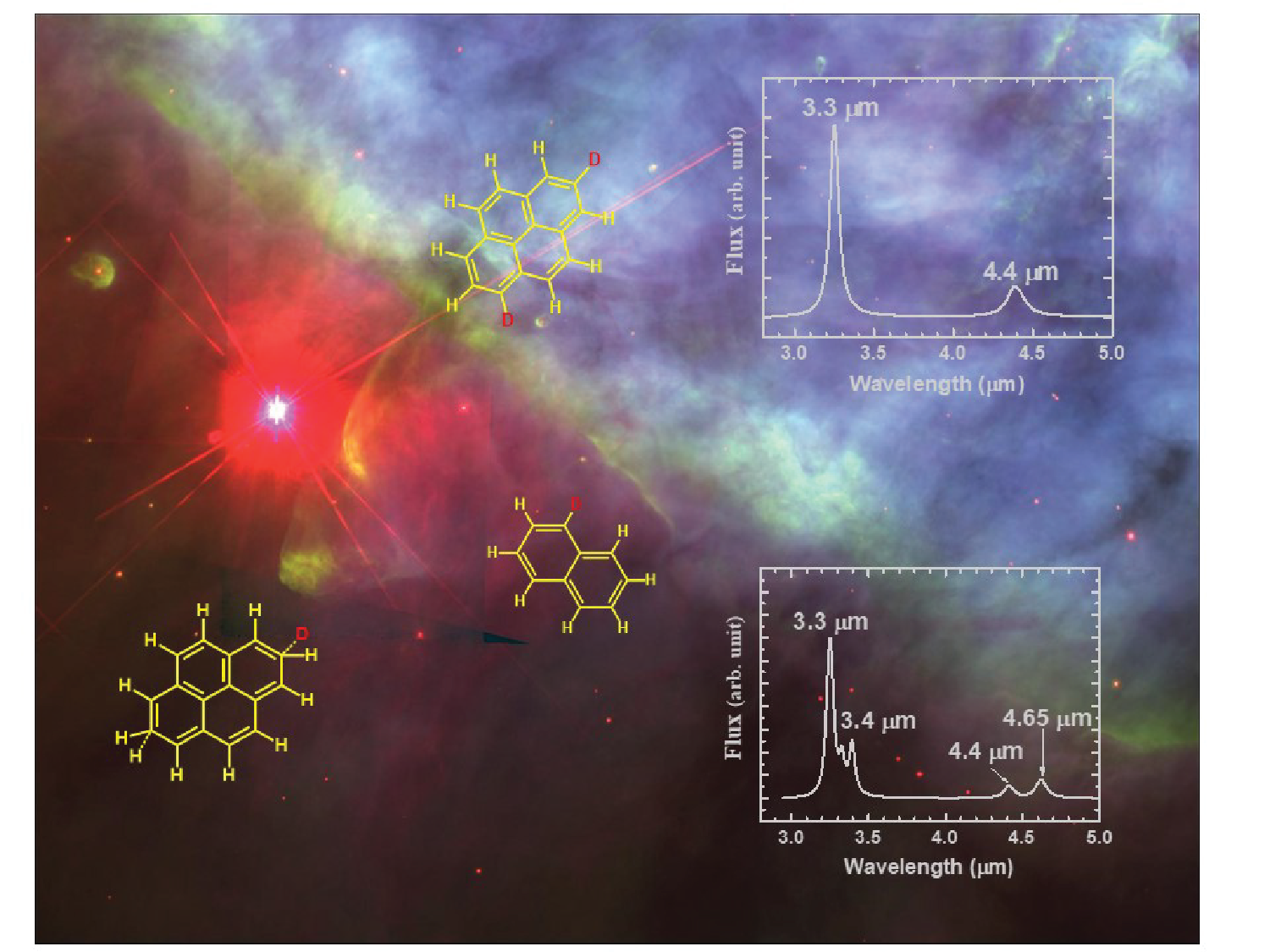}
}
\caption{\footnotesize
         \label{fig:route1}
         Schematic illustration of
         the deuteration of PAHs
         in the Orion Bar PDR,
         from regions near the ionization front
         where the C--D bonds are mostly aromatic,
         to regions intermediate beteen the ionization front
         and the dissociation front where PAHs
         could be superhydrogenated
         and even superdeuterated.
         }
\end{figure*}

As schematically illustrated in Figure~\ref{fig:route1},
it is natural to expect that interstellar PAHs
are deuterated in such regions.
Lets take the Orion Bar as an example.
The regions close to the ionization front
are rich in UV photons and deuteration could
occur during photodissociation events.
Upon absorption of an energetic photon,
one or more periphal H atoms could be stripped
off and replaced by D atoms. Due to the zero-point
energy difference, C--D bonds are more stable than
C--H bonds and therefore PAHs of intermediate size
are expected to become deuterium enriched
and obtain aromatic C--D units.
In these regions one expects to see the 4.4$\mum$
emission band arising from aromatic C--D stretch.

When moving away from the ionization front,
the regions intermediate beteen the ionization front
and the dissociation front, are rich in H and D atoms.
PAHs could be superhydrogenated
(i.e., two H atoms share one C atom)
and even superdeuterated
(i.e., one D atom and one H atom share one C atom).
In these regions one expects to see the aliphatic
C--D stretching emission at $\simali$4.6--4.8$\mum$
as well as the aliphatic C--H stretching emission
at 3.4$\mum$ (see Figure~\ref{fig:route1} for illustration).

\begin{figure*}
\centering{
  \includegraphics[scale=0.5,clip]{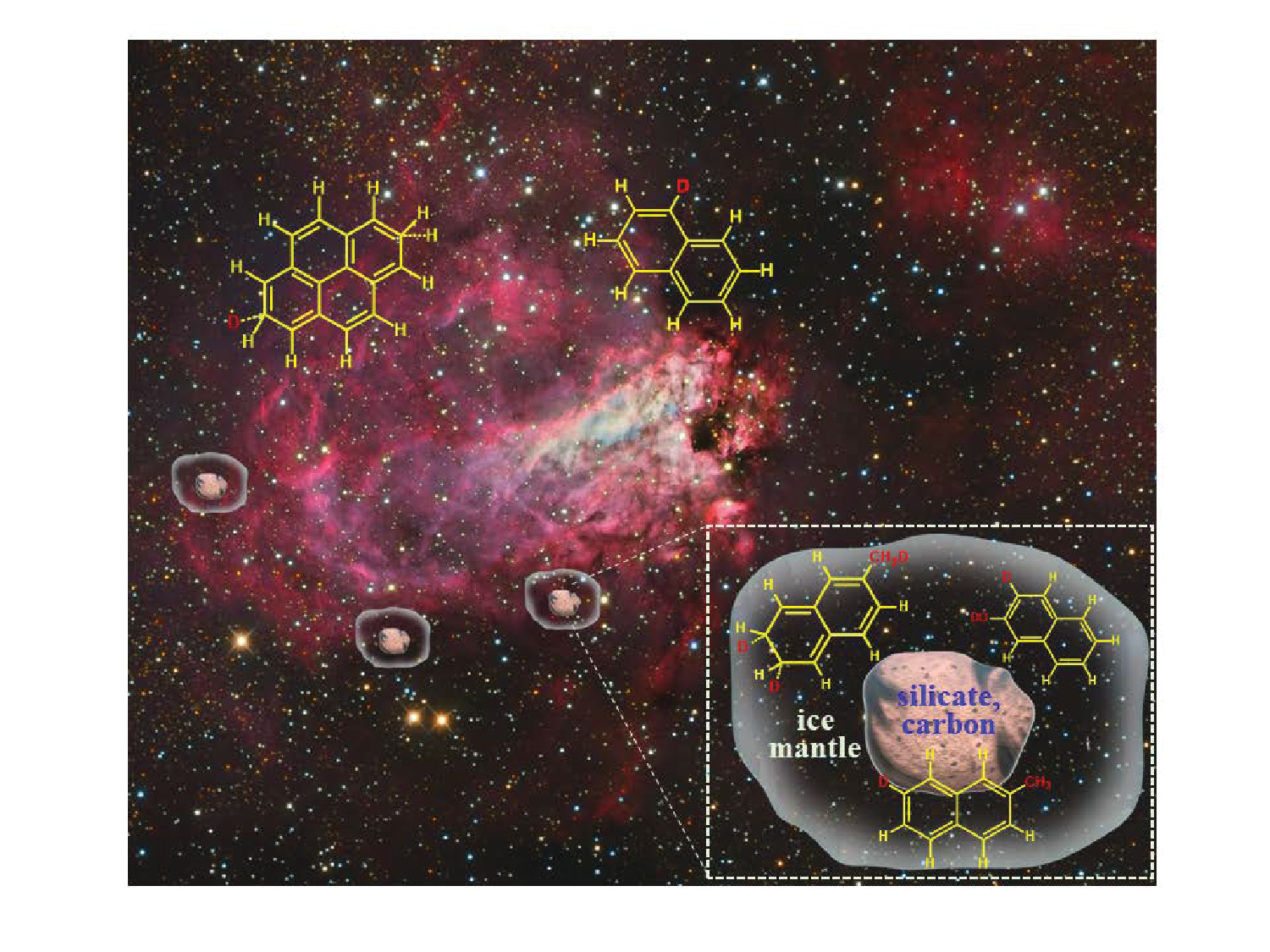}
}
\caption{\footnotesize
         \label{fig:route2}
         Schematic illustration of
         the deuteration of PAHs
         in the M17 dense molecular cloud.
         The deuteration occurs in the UV irradiated,
         D-enriched ice mantles coated on dust grains
         and generates aliphatic C--D units.
         }
\end{figure*}

Another way to deuterate PAHs and generate
aliphatic C--D bonds is through the addition
of an D-substituted methyl group.
This could occur in dense clouds
where PAHs and simple molecules
(e.g., CH$_4$) are frozen into the ice
mantles coated on dust grains.
In D-enriched ices, exposed to UV irradiation,
PAHs could attain an D-substituted methyl
group through chemical reactions with
deuterated methane
(see Figure~\ref{fig:route2}
for a schematic illustration).
PAHs could also become deuterated
in D-enriched ices by exchanging H atoms
with D atoms. This H/D exchange would
be enhanced for oxidized PAHs.
Furthermore, UV-stimulated addition
of D atoms at aromatic edge sites could
yield aliphatic C--D bonds.
Sandford et al.\ (2000) have experimentally
demonstrated that, when exposed to UV radiation,
PAHs in D-enriched ices could be oxidized
and easily deuterated.

\begin{figure*}
\centering{
\includegraphics[scale=0.3,clip]{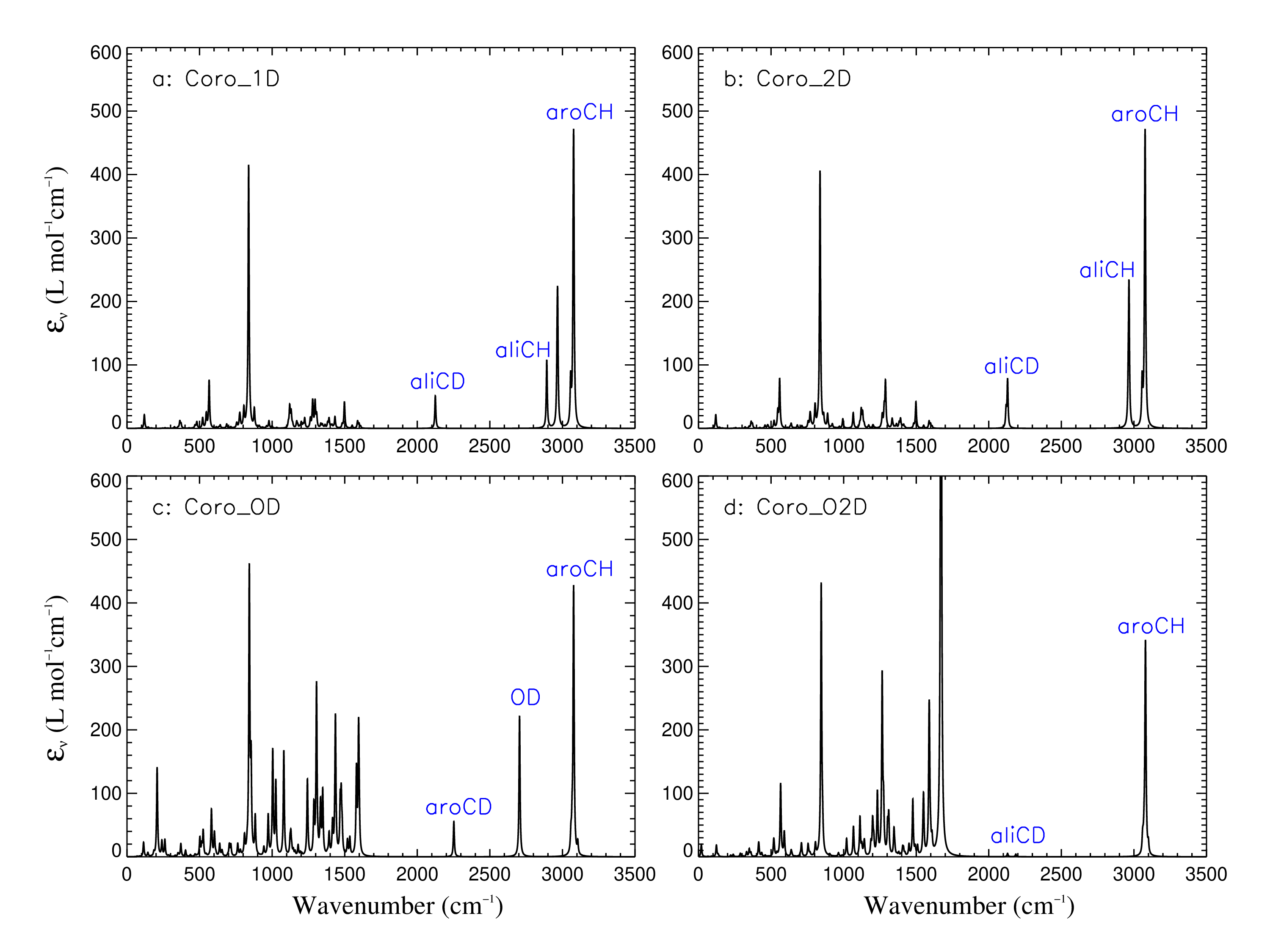}
}
\caption{\footnotesize
         \label{fig:Coro_HD_Spec}
         Vibrational spectra of ``superdeuterated'' coronene (a),
         doubly ``superdeuterated'' coronene (b),
         %(with two pairs of H and D atoms),
         and deuterated, oxidized coronenes (c, d).
         }
\end{figure*}

As deuterated PAHs made in UV-irradiated
ices are likely also oxidized, we also compute
the vibrational spectra of deuterated, oxidized
PAHs, taking coronene as an example.
We consider two coronene derivatives,
with D-atom exchange at oxidized edge sites
or at aromatic edge sites.
The structures of these two molecules
are also shown in Figure~\ref{fig:PAH_HD_Scheme}.
We show in Figure~\ref{fig:Coro_HD_Spec}
their vibrational spectra, together with two
``superdeuterated'' species.
The aliphatic and aromatic C--D stretches
as well as the O--D stretch are clearly seen.
We note that in Coro$\_$O2D
(in which two D atoms are attached
to one C atom)  the aliphatic C--D stretch
is significantly depressed.

\section{Summary}\label{sec:summary}
We have explored whether interstellar PAHs
could be a major reservoir of D to account for
its depletion from the gas phase,
focusing on the aliphatic C--D units of PAHs.
The major results are as follows:
\begin{enumerate}
\item We have employed the hybrid DFT method
          B3LYP in conjunction with the 6-311+G$^{\ast\ast}$
          basis set to calculate the vibrational spectra of
          deuterated PAHs with aliphatic C--D units,
          including ``superdeuterated'' PAHs in which
          one D and one H share an C atom
          and PAHs attached with an D-substituted
          methyl group.
          %...of various sizes, including benzene, naphthalene,
          %anthracene, phenanthrene, pyrene,
          %perylene and coronene.
%
\item For all these aliphatically deuterated species
          and their isomers, we have derived from
          the computed spectra
          the intrinsic band strengths of
          the 4.65$\mum$ aliphatic C--D stretch ($\Acd$),
          and the 3.3$\mum$ aromatic C--H stretch ($\ACH$).
          Both the C--H and C--D stretches predominantly
          arise from neutral PAHs.
          By averaging over all these molecules,
          we have determined the mean band strengths to be
          $\langle\Acd\rangle\approx15.3\pm3.3\km\mol^{-1}$,
          $\langle\Aaro\rangle\approx14.8\pm0.33\km\mol^{-1}$,
          and $\langle\Acd/\Aaro\rangle\approx1.04\pm0.21$.
\item We have compiled all the observational data
          from ISO/SWS and AKARI/IRC and derived
          $\left(I_{4.4}/I_{3.3}\right)_{\rm obs}$,
          the ratio of the power emitted from
          the aliphatic C--D band
          at 4.6--4.8$\mum$ ($I_{4.65}$)
          to that from the aromatic C--H band
          at 3.3$\mum$ ($I_{3.3}$).
          By comparing the computationally-derived mean ratio
          of $\langle\Acd/\Aaro\rangle\approx 1.04$ for
          aliphatically deuterated PAHs, it is found
          the degree of deuteration of some sources
          could be as high as $\simali$20\%.
          Combined with PAHs containing aromatic C--D units,
          the amount of D/H tied up in deuterated PAHs
          could exceed that missing from the gas phase,
          therefore consistent with the predictions
          from the BBN and GCE models
          and the gas-phase D/H of the local ISM.
\item     It is not clear how the high deuteration of PAHs
          required to account for the missing D is achieved.
          We have suggested that the deuteration of PAHs
          with aromatic C--D units may mainly occur
          in UV-rich regions, while the deuteration of
          PAHs with aliphatic C--D units may mainly
          occur in benign regions rich in H and D atoms
          or in dense clouds.
\end{enumerate}

\acknowledgments{%
We thank B.T.~Draine, A.N.~Witt
and the anonymous referee
for valuable suggestions.
XJY is supported in part by
NSFC 12122302 and 11873041.
AL is supported in part by NASA grants
80NSSC19K0572 and 80NSSC19K0701.
}

%%%%%%%%%%%%%%% References %%%%%%%%%%%%%%%%%%%%%%%%%%%%

%\clearpage

%=== Figures for Structures ===

%%% Figure 1 %%%
%%% Figure 1 %%%

%%% Figure 2 %%%
%%% Figure 2 %%%

%%% Figure 3 %%%
%%% Figure 3 %%%

%%% Figure 4 %%%
%%% Figure 4 %%%

%%% Figure 5 %%%
%%% Figure 5 %%%

% === Figures for Spectra: PAH_CH2D ===

%%% Figure 6 %%%
%%% Figure 6 %%%

%%% Figure 7 %%%
%%% Figure 7 %%%

% === Figures for Spectra: PAH_HD ===

%%% Figure 8 %%%
%%% Figure 8 %%%

%%% Figure 9 %%%
%%% Figure 9 %%%

% === Figures for Spectra: PAH_HD+ ===

%%% Figure 10 %%%
%%% Figure 10 %%%

%%% Figure 11 %%%
%%% Figure 11 %%%

\clearpage

% === Figures for Spectra: PAH_2D ===

%%% Figure 12 %%%
%%% Figure 12 %%%

%%% Figure 13 %%%
%%% Figure 13 %%%

%%% Figure 14 %%%
%%% Figure 14 %%%

%%% Figure 15 %%%
%%% Figure 15 %%%

%======Figures for Band Ratios======

%%% Figure 16 %%%
%%% Figure 16 %%%

%%% Figure 17 %%%
%%% Figure 17 %%%

%%% Figure 18 %%%
%%% Figure 18 %%%

%%% Figure 19 %%%
%%% Figure 19 %%%

%%% Figure 20 %%%
%%% Figure 20 %%%

%%% Figure 21 %%%
%%% Figure 21 %%%

%%% Figure 22 %%%
%%% Figure 22 %%%

%%% Figure 23 %%%
%%% Figure 23 %%%

%%% Figure 24  %%%
%%% Figure 24 %%%

\clearpage

%%%%% Table 1 %%%%%
\begin{table*}
\footnotesize
\begin{center}
\caption[]{\footnotesize
           Wavelengths ($\lambda$) and Intensities
           ($A_\lambda$ in $\km\mol^{-1}$)
           of the Nominal ``3.3$\mum$'' Aromatic C--H Stretch,
           ``3.4$\mum$'' Aliphatic C--H Stretch and
           ``4.65$\mum$'' Aliphatic C--D Stretch
           Computed at the B3LYP/6-311+G$^{\ast\ast}$ Level
           for All the Methyl-Deuterated PAHs
           Shown in Figure~\ref{fig:PAH_MethylD_Scheme}.
%           The Intensities $A_\lambda$ Relates to the Molar
%           Extinction Coefficients $\varepsilonWV$ through
%           Eq.\,\ref{eq:epsilon2A}.
           }
\label{tab:Freq_Int_PAH_MethylD}
\begin{tabular}{lccccccc}
\noalign{\smallskip} \hline \hline \noalign{\smallskip}
Compound	    &	$\lambda_{3.3}$   & $A_{3.3}$
                &	$\lambda_{3.4}$   & $A_{3.4}$
                &	$\lambda_{4.65}$  & $A_{4.65}$
                &   $\Aratio$          \\
                &  ($\mu$m)  &  (km$\mol^{-1}$)
                &  ($\mu$m)  & (km$\mol^{-1}$)
                &  ($\mu$m)  & (km$\mol^{-1}$)
                &                     \\
\noalign{\smallskip} \hline \noalign{\smallskip}
Benz$\_$CH$_2$D	&	3.25 	&	14.42 	&	3.37 	&	21.29 	&	4.60 	&	11.53 	&	0.80 	\\
Naph$\_$CH$_2$D	&	3.25 	&	13.88 	&	3.37 	&	22.58 	&	4.60 	&	12.37 	&	0.89 	\\
Anth$\_$CH$_2$D	&	3.25 	&	13.82 	&	3.37 	&	22.24 	&	4.59 	&	12.28 	&	0.89 	\\
Phen$\_$CH$_2$D	&	3.24 	&	13.07 	&	3.37 	&	22.64 	&	4.59 	&	12.43 	&	0.95 	\\
Pyre$\_$CH$_2$D	&	3.25 	&	14.66 	&	3.38 	&	23.93 	&	4.60 	&	13.62 	&	0.93 	\\
Pery$\_$CH$_2$D	&	3.25 	&	13.73 	&	3.38 	&	24.38 	&	4.60 	&	13.73 	&	1.00 	\\
Coro$\_$CH$_2$D	&	3.25 	&	16.58 	&	3.38 	&	24.83 	&	4.60 	&	14.24 	&	0.86 	\\\hline \noalign{\smallskip}
Average	&	3.25 	&	14.31 	&	3.37 	&	23.13 	&	4.60 	&	12.89 	&	0.90 	\\
stdev	&	0.00 	&	1.12 	&	0.00 	&	1.28 	&	0.00 	&	0.98 	&	0.07 	\\
\hline
\noalign{\smallskip} \noalign{\smallskip}
\end{tabular}
\end{center}
\end{table*}
%%%%% Table 1 %%%%%

%%%%% Table 2 %%%%%
\begin{table*}
\footnotesize
\begin{center}
\caption[]{\footnotesize
           Wavelengths ($\lambda$) and Intensities
           ($A_\lambda$ in $\km\mol^{-1}$)
           of the Nominal ``3.3$\mum$'' Aromatic C--H Stretch,
           ``3.4$\mum$'' Aliphatic C--H Stretch and
           ``4.65$\mum$'' Aliphatic C--D Stretch
           Computed at the B3LYP/6-311+G$^{\ast\ast}$ Level
           for All the ``Supedeuterated'' PAHs
           Shown in Figure~\ref{fig:PAH_HD_Scheme}.
%           The Intensities $A_\lambda$ Relates to the Molar
%           Extinction Coefficients $\varepsilonWV$ through
%           Eq.\,\ref{eq:epsilon2A}.
           }
\label{tab:Freq_Int_PAH_HD}
\begin{tabular}{lccccccc}
\noalign{\smallskip} \hline \hline \noalign{\smallskip}
Compound	    &	$\lambda_{3.3}$ & $A_{3.3}$
                &	$\lambda_{3.4}$ & $A_{3.4}$
                &	$\lambda_{4.65}$  & $A_{4.65}$
                &   $\Aratio$   \\
                &  ($\mu$m)  &  (km$\mol^{-1}$)
                &  ($\mu$m)  & (km$\mol^{-1}$)
                &  ($\mu$m)  & (km$\mol^{-1}$)
                & \\
\noalign{\smallskip} \hline \noalign{\smallskip}
Benz$\_$HD	&	3.26 	&	16.47 	&	3.43 	&	32.84 	&	4.67 	&	17.82 	&	1.08 	\\
Naph$\_$HD	&	3.25 	&	14.69 	&	3.42 	&	32.39 	&	4.70 	&	19.32 	&	1.31 	\\
Anth$\_$HD	&	3.25 	&	14.47 	&	3.43 	&	28.88 	&	4.66 	&	15.98 	&	1.10 	\\
Phen$\_$HD	&	3.25 	&	13.59 	&	3.42 	&	32.29 	&	4.66 	&	17.82 	&	1.31 	\\
Pyre$\_$HD	&	3.25 	&	15.07 	&	3.42 	&	32.18 	&	4.62 	&	17.69 	&	1.17 	\\
Pery$\_$HD	&	3.24 	&	13.71 	&	3.43 	&	33.05 	&	4.65 	&	17.82 	&	1.30 	\\
Coro$\_$HD	&	3.27 	&	17.24 	&	3.40 	&	37.89 	&	4.71 	&	6.97 	&	0.40 	\\
\hline \noalign{\smallskip}
Average	&	3.25 	&	15.03 	&	3.42 	&	32.79 	&	4.66 	&	16.20 	&	1.10 	\\
Stdev	&	0.01 	&	1.36 	&	0.01 	&	2.65 	&	0.03 	&	4.18 	&	0.32 	\\
\hline
\noalign{\smallskip} \noalign{\smallskip}
\end{tabular}
\end{center}
\end{table*}
%%%%% Table 2 %%%%%

%%%%% Table 3 %%%%%
\begin{table*}
\footnotesize
\begin{center}
\caption[]{\footnotesize
           Wavelengths ($\lambda$) and Intensities
           ($A_\lambda$ in $\km\mol^{-1}$)
           of the Nominal ``3.3$\mum$'' Aromatic C--H Stretch,
           ``3.4$\mum$'' Aliphatic C--H Stretch
           ``4.65$\mum$'' Aliphatic C--D Stretch
           Computed at the B3LYP/6-311+G$^{\ast\ast}$ Level
           for All the ``Superdeuterated'' PAH Cations
           Shown in Figure~\ref{fig:PAH_HDPlus_Scheme}.
%           The Intensities $A_\lambda$ Relates to the Molar
%           Extinction Coefficients $\varepsilonWV$ through
%           Eq.\,\ref{eq:epsilon2A}.
           }
\label{tab:Freq_Int_PAH_HDPlus}
\begin{tabular}{lccccccc}
\noalign{\smallskip} \hline \hline \noalign{\smallskip}
Compound	    &	$\lambda_{3.3}$ & $A_{3.3}$
                &	$\lambda_{3.4}$ & $A_{3.4}$
                &	$\lambda_{4.65}$  & $A_{4.65}$  &  $\Aratio$  \\
                &  ($\mu$m)  &  (km$\mol^{-1}$)
                &  ($\mu$m)  & (km$\mol^{-1}$)
                &  ($\mu$m)  & (km$\mol^{-1}$)    &               \\
\noalign{\smallskip} \hline \noalign{\smallskip}
Benz$\_$HD+	&	3.24 	&	9.01 	&	3.31 	&	0.13 	&	4.81 	&	1.61 	&	0.18 	\\
Naph$\_$HD+	&	3.22 	&	0.84 	&	3.51 	&	0.22 	&	4.66 	&	0.64 	&	0.76 	\\
Anth$\_$HD+	&	3.24 	&	1.32 	&	3.51 	&	2.28 	&	4.79 	&	2.23 	&	1.68 	\\
Phen$\_$HD+	&	3.23 	&	0.33 	&	3.49 	&	2.15 	&	4.75 	&	1.18 	&	3.56 	\\
Pyre$\_$HD+	&	3.24 	&	0.46 	&	3.50 	&	1.39 	&	4.78 	&	0.95 	&	2.09 	\\
Pery$\_$HD+	&	3.22 	&	1.10 	&	3.51 	&	2.74 	&	4.78 	&	2.27 	&	2.07 	\\
Coro$\_$HD+	&	3.23 	&	1.92 	&	3.48 	&	5.20 	&	4.75 	&	2.62 	&	1.36 	\\
\hline \noalign{\smallskip}
Average	&	3.23 	&	2.14 	&	3.47 	&	2.02 	&	4.76 	&	1.64 	&	1.67 	\\
Stdev	&	0.01 	&	3.08 	&	0.07 	&	1.73 	&	0.05 	&	0.75 	&	1.00 	\\
\hline
\noalign{\smallskip} \noalign{\smallskip}
\end{tabular}
\end{center}
\end{table*}
%%%%% Table 3 %%%%%

%%%%% Table 4 %%%%%
\begin{table*}
\footnotesize
\begin{center}
\caption[]{\footnotesize
           Wavelengths ($\lambda$) and Intensities
           ($A_\lambda$ in $\km\mol^{-1}$)
           of the Nominal ``3.3$\mum$'' Aromatic C--H Stretch,
           ``3.4$\mum$'' Aliphatic C--H Stretch,
           ``4.4$\mum$'' Aromatic C--D Stretch and
           ``4.65$\mum$'' Aliphatic C--D Stretch
           Computed at the B3LYP/6-311+G$^{\ast\ast}$ Level
           for Deuterated Pyrenes
           Containing a Methyl-Deuterated Sidegroup
           as well as a Periphal D Atom
           Shown in Figure~\ref{fig:PAH_Methyl2D_Scheme}.
%           The Intensities $A_\lambda$ Relates to the Molar
%           Extinction Coefficients $\varepsilonWV$ through
%           Eq.\,\ref{eq:epsilon2A}.
           }
\label{tab:Freq_Int_PAH_Methyl2D}
\begin{tabular}{lccccccccc}
\noalign{\smallskip} \hline \hline \noalign{\smallskip}
Compound	    &	$\lambda_{3.3}$   & $A_{3.3}$
                &	$\lambda_{3.4}$   & $A_{3.4}$
                &	$\lambda_{4.4}$   & $A_{4.4}$
                &	$\lambda_{4.65}$  & $A_{4.65}$
                &   $\Aratio$                   \\
                &  ($\mu$m)  &  (km$\mol^{-1}$)
                &  ($\mu$m)  & (km$\mol^{-1}$)
                &  ($\mu$m)  & (km$\mol^{-1}$)
                &  ($\mu$m)  & (km$\mol^{-1}$)
                &                               \\
\noalign{\smallskip} \hline \noalign{\smallskip}
Pyre1$\_$D1	&	3.25 	&	15.30 	&	3.37 	&	26.61 	&	4.41 	&	6.52 	&	4.62 	&	11.63 	&	0.76 	\\
Pyre1$\_$D2	&	3.25 	&	15.79 	&	3.37 	&	26.75 	&	4.38 	&	4.15 	&	4.62 	&	11.51 	&	0.73 	\\
Pyre1$\_$D3	&	3.25 	&	15.16 	&	3.37 	&	26.57 	&	4.41 	&	7.46 	&	4.62 	&	11.64 	&	0.77 	\\
Pyre1$\_$D4	&	3.25 	&	15.10 	&	3.37 	&	26.61 	&	4.41 	&	7.67 	&	4.62 	&	11.63 	&	0.77 	\\
Pyre1$\_$D5	&	3.25 	&	13.54 	&	3.37 	&	27.97 	&	4.42 	&	13.25 	&	4.62 	&	11.65 	&	0.86 	\\
Pyre1$\_$D6	&	3.25 	&	15.25 	&	3.37 	&	26.56 	&	4.41 	&	7.11 	&	4.62 	&	11.62 	&	0.76 	\\
Pyre1$\_$D7	&	3.25 	&	15.21 	&	3.37 	&	26.61 	&	4.41 	&	6.71 	&	4.62 	&	11.63 	&	0.76 	\\
Pyre1$\_$D8	&	3.25 	&	14.25 	&	3.37 	&	26.59 	&	4.40 	&	12.23 	&	4.62 	&	11.63 	&	0.82 	\\
Pyre1$\_$D9	&	3.25 	&	15.34 	&	3.37 	&	26.61 	&	4.41 	&	6.31 	&	4.62 	&	11.63 	&	0.76 	\\ \hline \noalign{\smallskip}
Pyre2$\_$D1	&	3.26 	&	14.62 	&	3.38 	&	23.99 	&	4.41 	&	6.65 	&	4.57 	&	15.39 	&	1.05 	\\
Pyre2$\_$D4	&	3.26 	&	14.56 	&	3.38 	&	24.00 	&	4.41 	&	6.96 	&	4.57 	&	15.35 	&	1.05 	\\
Pyre2$\_$D5	&	3.25 	&	14.40 	&	3.38 	&	24.02 	&	4.43 	&	7.72 	&	4.57 	&	14.91 	&	1.04 	\\
Pyre2$\_$D6	&	3.26 	&	14.51 	&	3.38 	&	23.99 	&	4.41 	&	7.22 	&	4.57 	&	15.39 	&	1.06 	\\
Pyre2$\_$D7	&	3.26 	&	13.65 	&	3.38 	&	23.99 	&	4.40 	&	11.60 	&	4.57 	&	15.39 	&	1.13 	\\ \hline \noalign{\smallskip}
Pyre4$\_$D1	&	3.25 	&	15.05 	&	3.37 	&	22.84 	&	4.41 	&	6.26 	&	4.62 	&	11.67 	&	0.78 	\\
Pyre4$\_$D2	&	3.25 	&	14.88 	&	3.37 	&	22.81 	&	4.41 	&	7.19 	&	4.62 	&	11.67 	&	0.78 	\\
Pyre4$\_$D3	&	3.25 	&	14.87 	&	3.37 	&	22.82 	&	4.41 	&	7.32 	&	4.62 	&	11.68 	&	0.79 	\\
Pyre4$\_$D4	&	3.25 	&	15.68 	&	3.37 	&	22.90 	&	4.37 	&	3.36 	&	4.62 	&	11.60 	&	0.74 	\\
Pyre4$\_$D5	&	3.25 	&	14.10 	&	3.37 	&	24.13 	&	4.43 	&	9.22 	&	4.62 	&	11.67 	&	0.83 	\\
Pyre4$\_$D6	&	3.25 	&	14.00 	&	3.37 	&	22.82 	&	4.40 	&	12.06 	&	4.62 	&	11.68 	&	0.83 	\\
Pyre4$\_$D7	&	3.25 	&	15.05 	&	3.37 	&	22.83 	&	4.41 	&	6.09 	&	4.62 	&	11.67 	&	0.78 	\\
Pyre4$\_$D8	&	3.25 	&	14.99 	&	3.37 	&	22.81 	&	4.41 	&	6.66 	&	4.62 	&	11.67 	&	0.78 	\\
Pyre4$\_$D9	&	3.25 	&	14.07 	&	3.37 	&	22.78 	&	4.40 	&	11.93 	&	4.62 	&	11.69 	&	0.83 	\\\hline \noalign{\smallskip}
average	&	3.25 	&	14.76 	&	3.37 	&	24.68 	&	4.41 	&	7.90 	&	4.61 	&	12.43 	&	0.85 	\\
stdev 	&	0.00 	&	0.61 	&	0.00 	&	1.80 	&	0.01 	&	2.60 	&	0.02 	&	1.54 	&	0.12 	\\
\hline
\noalign{\smallskip} \noalign{\smallskip}
\end{tabular}
\end{center}
\end{table*}
%%%%% Table 4 %%%%%

%%%%% Table 5 %%%%%
\begin{table*}
\footnotesize
\begin{center}
\caption[]{\footnotesize
           Wavelengths ($\lambda$) and Intensities
           ($A_\lambda$ in $\km\mol^{-1}$)
           of the Nominal ``3.3$\mum$'' Aromatic C--H Stretch,
           ``3.4$\mum$'' Aliphatic C--H Stretch,
           ``4.4$\mum$'' Aromatic C--D Stretch and
           ``4.65$\mum$'' Aliphatic C--D Stretch
           Computed at the B3LYP/6-311+G$^{\ast\ast}$ Level
           for All the ``Superdeuterated'' Pyrenes
           Containing an H+H Pair, an H+D Pair,
           as well as a Periphal D Atom.
           Shown in Figure~\ref{fig:PAH_H2D_Scheme}.
%           The Intensities $A_\lambda$ Relates to the Molar
%           Extinction Coefficients $\varepsilonWV$ through
%           Eq.\,\ref{eq:epsilon2A}.
           }
\label{tab:Freq_Int_PAH_H2D}
\begin{tabular}{lccccccccc}
\noalign{\smallskip} \hline \hline \noalign{\smallskip}
Compound	    &	$\lambda_{3.3}$   & $A_{3.3}$
                &	$\lambda_{3.4}$   & $A_{3.4}$
                &	$\lambda_{4.4}$   & $A_{4.4}$
                &	$\lambda_{4.65}$  & $A_{4.65}$ & $\Aratio$  \\
                &  ($\mu$m)  &  (km$\mol^{-1}$)
                &  ($\mu$m)  & (km$\mol^{-1}$)
                &  ($\mu$m)  & (km$\mol^{-1}$)
                &  ($\mu$m)  & (km$\mol^{-1}$)     &            \\
\noalign{\smallskip} \hline \noalign{\smallskip}
Pyre1D$\_$D1	&	3.26 	&	15.29 	&	3.44 	&	37.86 	&	4.41 	&	5.17 	&	4.59 	&	18.00 	&	1.18 	\\
Pyre1D$\_$D2	&	3.26 	&	15.06 	&	3.44 	&	37.86 	&	4.41 	&	5.66 	&	4.59 	&	18.02 	&	1.20 	\\
Pyre1D$\_$D3	&	3.26 	&	14.65 	&	3.44 	&	37.86 	&	4.41 	&	7.96 	&	4.59 	&	18.00 	&	1.23 	\\
Pyre1D$\_$D4	&	3.25 	&	14.43 	&	3.44 	&	37.89 	&	4.42 	&	9.79 	&	4.59 	&	17.47 	&	1.21 	\\
Pyre1D$\_$D5	&	3.26 	&	15.13 	&	3.44 	&	37.86 	&	4.41 	&	6.31 	&	4.59 	&	18.03 	&	1.19 	\\
Pyre1D$\_$D6	&	3.26 	&	13.70 	&	3.44 	&	37.85 	&	4.40 	&	13.22 	&	4.59 	&	18.00 	&	1.31 	\\
Pyre1D$\_$D7	&	3.25 	&	14.76 	&	3.44 	&	37.86 	&	4.41 	&	8.09 	&	4.59 	&	18.01 	&	1.22 	\\
Pyre1D$\_$D8	&	3.25 	&	14.97 	&	3.44 	&	38.19 	&	4.43 	&	5.01 	&	4.59 	&	18.23 	&	1.22 	\\
\hline\noalign{\smallskip}
Pyre2D$\_$D1	&	3.26 	&	15.28 	&	3.45 	&	35.10 	&	4.41 	&	5.16 	&	4.60 	&	22.58 	&	1.48 	\\
Pyre2D$\_$D2	&	3.26 	&	15.05 	&	3.45 	&	35.09 	&	4.41 	&	5.67 	&	4.60 	&	22.59 	&	1.50 	\\
Pyre2D$\_$D3	&	3.26 	&	14.65 	&	3.45 	&	35.09 	&	4.41 	&	7.96 	&	4.60 	&	22.58 	&	1.54 	\\
Pyre2D$\_$D4	&	3.25 	&	14.28 	&	3.44 	&	35.52 	&	4.42 	&	9.06 	&	4.60 	&	22.60 	&	1.58 	\\
Pyre2D$\_$D5	&	3.26 	&	15.13 	&	3.45 	&	35.09 	&	4.41 	&	6.35 	&	4.60 	&	22.58 	&	1.49 	\\
Pyre2D$\_$D6	&	3.26 	&	13.69 	&	3.45 	&	35.09 	&	4.40 	&	13.21 	&	4.60 	&	22.59 	&	1.65 	\\
Pyre2D$\_$D7	&	3.25 	&	14.75 	&	3.45 	&	35.09 	&	4.41 	&	8.10 	&	4.60 	&	22.58 	&	1.53 	\\
Pyre2D$\_$D8	&	3.25 	&	15.11 	&	3.45 	&	35.05 	&	4.43 	&	5.61 	&	4.60 	&	22.25 	&	1.47 	\\
\hline\noalign{\smallskip}
Pyre4D$\_$D1	&	3.25 	&	15.20 	&	3.41 	&	26.54 	&	4.42 	&	7.47 	&	4.59 	&	18.20 	&	1.20 	\\
Pyre4D$\_$D2	&	3.25 	&	15.65 	&	3.41 	&	26.08 	&	4.41 	&	7.01 	&	4.59 	&	18.16 	&	1.16 	\\
Pyre4D$\_$D3	&	3.25 	&	15.64 	&	3.41 	&	26.09 	&	4.41 	&	7.00 	&	4.59 	&	18.17 	&	1.16 	\\
Pyre4D$\_$D4	&	3.25 	&	15.11 	&	3.41 	&	26.08 	&	4.40 	&	8.74 	&	4.59 	&	18.16 	&	1.20 	\\
Pyre4D$\_$D5	&	3.25 	&	15.47 	&	3.41 	&	26.09 	&	4.41 	&	8.00 	&	4.59 	&	18.16 	&	1.17 	\\
Pyre4D$\_$D6	&	3.25 	&	15.48 	&	3.41 	&	26.08 	&	4.41 	&	8.01 	&	4.59 	&	18.16 	&	1.17 	\\
Pyre4D$\_$D7	&	3.25 	&	15.11 	&	3.41 	&	26.09 	&	4.40 	&	8.71 	&	4.59 	&	18.17 	&	1.20 	\\
Pyre4D$\_$D8	&	3.25 	&	15.36 	&	3.41 	&	26.12 	&	4.42 	&	8.23 	&	4.59 	&	17.64 	&	1.15 	\\
\hline \noalign{\smallskip}
Average	&	3.25 	&	14.96 	&	3.43 	&	33.06 	&	4.41 	&	7.73 	&	4.59 	&	19.54 	&	1.31 	\\
Stdev	&	0.00 	&	0.52 	&	0.02 	&	5.13 	&	0.01 	&	2.16 	&	0.00 	&	2.18 	&	0.17 	\\
\hline
\noalign{\smallskip} \noalign{\smallskip}
\end{tabular}
\end{center}
\end{table*}
%%%%% Table 5 %%%%%

%%%%% Table 6 %%%%%
\begin{table*}
%\tiny
\footnotesize
\begin{center}
\caption[]{\footnotesize
           {\it Mean} Intensities of
           the 3.3$\mum$ Aromatic C--H Stretch ($\ACH$) and
           the 4.65$\mum$ Aliphatic C--D Stretch ($\Acd$)
           Computed at the {\rm B3LYP/6-311+G$^{\ast\ast}$} Level
           for All the Deuterated PAHs Considered
           in This Work.
           The $\ACH$ and $\Acd$
           Band Strengths Are on a Per C--H or C--D Bond Basis.
           %Also Tabulated Are the Recommended Band Strengths
           %Obtained by Averaging over Mono-Deuterated PAHs
           %and Multi-Deuterated Pyrenes.
           }
\label{tab:MeanBandStrengths}
\begin{tabular}{lcc}
%\specialrule{0em}{1pt}{1pt}
%\noalign{\smallskip} \hline \hline \noalign{\smallskip}
\noalign{\smallskip}\noalign{\smallskip} \hline\hline
Band Strengths	    &	Neutral  PAHs  & Cationic PAHs \\\hline
%\noalign{\smallskip} \hline \noalign{\smallskip}
      $A_{3.3} $ (km$\mol^{-1}$)     &  14.76$\pm{0.33}$       &    2.14$^{+3.08}_{-2.14}$   \\
      $A_{4.65}$ (km$\mol^{-1}$)     &  15.26$\pm{3.31}$       &    1.64$\pm{0.75}$          \\
      $\Aratio$                      &  1.04$\pm{0.21}$        &    1.67$\pm{1.00}$          \\\hline
\noalign{\smallskip} \noalign{\smallskip}
\end{tabular}
\end{center}
\end{table*}
%%%%% Table 6  %%%%%

%%%%% Table 7 %%%%%
\begin{sidewaystable}[h]
%\begin{table*}
\tiny
%\scriptsize
\begin{center}
\caption[]{\footnotesize
                A Summary of the Power Emitted
                from the 3.3$\mum$ Aromatic
                C--H Stretch $\left(I_{3.3}\right)_{\rm obs}$,
                the 3.4$\mum$ Aliphatic
                C--H Stretch $\left(I_{3.4}\right)_{\rm obs}$,
                the 4.4$\mum$ Aromatic
                C--D Stretch $\left(I_{4.4}\right)_{\rm obs}$, and
                the 4.65$\mum$ Aliphatic
                C--D Stretch $\left(I_{4.65}\right)_{\rm obs}$
                Compiled from Astronomical Observations.
                Also Shown Are the Degree of Deuteration
                Derived for Each Source
                }
\label{tab:I465I33_obs}
\begin{tabular}{lcccccccccccccc}
\noalign{\smallskip} \hline \hline \noalign{\smallskip}
Object	    &  Type   &  	\multicolumn{1}{c}{aromatic C--H}  &
     & \multicolumn{3}{c}{aliphatic C--H} &  & \multicolumn{3}{c}{aromatic C--D}     &   &   \multicolumn{3}{c}{aliphatic C--D}
\\ \noalign{\smallskip}    \cline{5-7}      \cline{9-11} \cline{13-15} \noalign{\smallskip}
 	    &      &  	\multicolumn{1}{c}{$\left(I_{3.3}\right)_{\rm obs}^\dag$}  &
 & \multicolumn{1}{c}{$\left(I_{3.4}\right)_{\rm obs}^\dag$}  & \multicolumn{1}{c}{$\left(I_{3.4}/I_{3.3}\right)_{\rm obs}^\dag$}  &  $\faliCH$ &
 &  \multicolumn{1}{c}{$\left(I_{4.4}\right)_{\rm obs}^\dag$} & \multicolumn{1}{c}{$\left(I_{4.4}/I_{3.3}\right)_{\rm obs}^\dag$}  &  $\faroCD$ &
 & \multicolumn{1}{c}{$\left(I_{4.65}\right)_{\rm obs}^\dag$} & \multicolumn{1}{c}{$\left(I_{4.65}/I_{3.3}\right)_{\rm obs}^\dag$}  &  $\faliCD$
\\\noalign{\smallskip} \hline \noalign{\smallskip}
G75.78+0.34$^a$	&	 H{\sc ii} Region	&	22.2	$\pm{	2.0	}$&&	13.3 	$\pm{	2.1 	}$&	59.9	$\pm{	4.1	}$&	34.0	$\pm{	4.7	}$&&	0.30	$\pm{	0.19	}$&	1.35	$\pm{	0.73	}$&	3.3	$\pm{	1.3	}$&&	2.25	$\pm{	0.86	}$&	10.1	$\pm{	2.96	}$&	7.1	$\pm{	4.1	}$\\
NGC3603$^a$	&	 H{\sc ii} Region	&	25.1	$\pm{	2.4	}$&&	16.7 	$\pm{	3.1 	}$&	66.5	$\pm{	6.0	}$&	37.8	$\pm{	5.7	}$&&	0.62	$\pm{	0.52	}$&	2.47	$\pm{	1.84	}$&	5.9	$\pm{	2.8	}$&&	8.70	$\pm{	4.41	}$&	34.7	$\pm{	14.3	}$&	20.6	$\pm{	16.0	}$\\
W51 obs.1$^a$	&	 H{\sc ii} Region	&	20.0	$\pm{	1.8	}$&&	11.2 	$\pm{	1.5 	}$&	56.0	$\pm{	2.5	}$&	31.8	$\pm{	4.0	}$&&	0.41	$\pm{	0.14	}$&	2.05	$\pm{	0.52	}$&	5.0	$\pm{	1.5	}$&&	0.88	$\pm{	0.47	}$&	4.40	$\pm{	1.95	}$&	3.2	$\pm{	2.1	}$\\
W51 obs.2$^a$	&	 H{\sc ii} Region	&	20.6	$\pm{	1.8	}$&&	11.5 	$\pm{	1.4 	}$&	55.8	$\pm{	1.9	}$&	31.7	$\pm{	3.9	}$&&	0.16	$\pm{	0.14	}$&	0.78	$\pm{	0.61	}$&	1.9	$\pm{	0.9	}$&&	0.86	$\pm{	0.47	}$&	4.17	$\pm{	1.92	}$&	3.0	$\pm{	2.0	}$\\
M8$^a$	&	 H{\sc ii} Region	&	99.1	$\pm{	8.9	}$&&	55.2 	$\pm{	6.8 	}$&	55.7	$\pm{	1.9	}$&	31.6	$\pm{	3.9	}$&&	1.03	$\pm{	0.53	}$&	1.04	$\pm{	0.44	}$&	2.6	$\pm{	0.9	}$&&	1.87	$\pm{	1.63	}$&	1.89	$\pm{	1.48	}$&	1.4	$\pm{	1.3	}$\\
IRAS112073 obs.1$^a$	&	 H{\sc ii} Region	&	11.7	$\pm{	1.0	}$&&	6.7 	$\pm{	0.8 	}$&	57.3	$\pm{	1.9	}$&	32.5	$\pm{	4.0	}$&&	0.30	$\pm{	0.08	}$&	2.56	$\pm{	0.46	}$&	6.1	$\pm{	1.8	}$&&	4.40	$\pm{	0.76	}$&	37.6	$\pm{	3.28	}$&	22.0	$\pm{	13.1	}$\\
IRAS112073 obs.2$^a$	&	 H{\sc ii} Region	&	12.2	$\pm{	1.0	}$&&	6.1 	$\pm{	0.7 	}$&	50.0	$\pm{	1.6	}$&	28.4	$\pm{	3.5	}$&&	0.37	$\pm{	0.08	}$&	3.03	$\pm{	0.41	}$&	7.2	$\pm{	2.1	}$&&	2.92	$\pm{	0.76	}$&	23.9	$\pm{	4.27	}$&	15.2	$\pm{	8.8	}$\\
M17b$^a$	&	PDR	&	12.3	$\pm{	1.5	}$&&	6.6 	$\pm{	1.8 	}$&	53.7	$\pm{	8.1	}$&	30.5	$\pm{	5.9	}$&&	-		-	&	-		-	&	-		-	&&	3.32	$\pm{	1.94	}$&	27.0	$\pm{	12.5	}$&	16.8	$\pm{	13.2	}$\\
M17$^b$	&	PDR	&	14.9	$\pm{	0.2	}$&&	4.8 	$\pm{	0.1 	}$&	32.2	$\pm{	0.2	}$&	18.3	$\pm{	2.2	}$&&	0.14	$\pm{	0.04	}$&	0.94	$\pm{	0.26	}$&	2.3	$\pm{	0.7	}$&&	0.32	$\pm{	0.06	}$&	2.15	$\pm{	0.37	}$&	1.6	$\pm{	0.8	}$\\
Orion Bar$^b$	&	PDR	&	53.1	$\pm{	0.6	}$&&	19.2 	$\pm{	0.3 	}$&	36.2	$\pm{	0.2	}$&	20.5	$\pm{	2.5	}$&&	1.36	$\pm{	0.11	}$&	2.56	$\pm{	0.18	}$&	6.1	$\pm{	1.7	}$&&	0.76	$\pm{	0.13	}$&	1.43	$\pm{	0.23	}$&	1.1	$\pm{	0.5	}$\\
G18.14.0$^b$	&	Reflection Nebula	&	4.78	$\pm{	0.06	}$&&	1.99 	$\pm{	0.04 	}$&	41.6	$\pm{	0.3	}$&	23.7	$\pm{	2.8	}$&&	0.12	$\pm{	0.01	}$&	2.45	$\pm{	0.18	}$&	5.9	$\pm{	1.6	}$&&	0.03	$\pm{	0.01	}$&	0.56	$\pm{	0.20	}$&	0.4	$\pm{	0.2	}$\\

\hline \noalign{\smallskip}
\end{tabular}
\\
$^\dag$ Flux densities in units of  $10^{-17}$\,W\,m$^{-2}$\,arcsec$^{-2}$     \\
$^a$  Doney et al.\ 2016          \\
$^b$  Onaka et al.\ 2014          \\

\end{center}
%\end{table*}
\end{sidewaystable}
%%%%% Table 5 %%%%%

\end{document}